\documentclass[11pt,tightenlines,nofootinbib,preprintnumbers,showpacs,titlepage,eqsecnum,prd]{revtex4}

\usepackage{slashed}
\usepackage{amsmath}
\usepackage{amssymb}
\usepackage{axodraw}
\usepackage{graphicx}

\providecommand{\openone}{\leavevmode\hbox{\large1\kern-7.3pt\normalsize1}}
\newcommand{\be}{\begin{equation}}
\newcommand{\ee}{\end{equation}}
\newcommand{\ba}{\begin{eqnarray}}
\newcommand{\ea}{\end{eqnarray}}
\newcommand{\rmi}[1]{{\mbox{\tiny #1}}}

\newcommand{\nr}[1]{(\ref{#1})}
\newcommand{\tr}{{\rm Tr\,}}
\newcommand{\nn}{\nonumber \\}
\newcommand{\fr}[2]{{\frac{#1}{#2}\,}}
\newcommand{\msbar}{{\overline{\mbox{\rm MS}}}}
\newcommand{\lambdamsbar}{{\Lambda_{\overline {\mbox{\tiny MS}} }}}

\renewcommand{\(}{\left(}
\renewcommand{\)}{\right)}

\renewcommand{\d}{\delta}
\newcommand{\e}{\epsilon}

\newcommand{\0}{\over }

\newcommand{\tinyMSbar}{{\overline{\mbox{\tiny\rm{MS}}}}}

\def\sumint{\hbox{$\sum$}\!\!\!\!\!\!\!\int}
\newcommand{\I}{{\cal I}^0}
\newcommand{\It}{\widetilde{\cal I}^0}

\renewcommand{\ln}{{\rm ln}}
\newcommand{\mubar}{\bar{\mu}}
\newcommand{\zb}{\frac{\zeta'(-3)}{\zeta(-3)}}
\newcommand{\za}{\frac{\zeta'(-1)}{\zeta(-1)}}

\newcommand{\imathb}{i}

\newcommand{\mmag}{m_\rmi{mag}}
\newcommand{\mmagb}{\overline{m}_\rmi{mag}}
\newcommand{\im}{{\rm Im}\,}
\newcommand{\re}{{\rm Re}\,}

\newcommand{\pic}[1]{\;\parbox[c]{45pt}{\begin{picture}(45,30)(0,0)
\SetWidth{1.0}\SetScale{1.0} #1 \end{picture}}\;}

\newcommand{\picb}[1]{\;\parbox[c]{48pt}{\begin{picture}(45,30)(-9,0)
\SetWidth{1.0}\SetScale{1.0} #1 \end{picture}}\;}
\newcommand{\picc}[1]{\;\parbox[c]{45pt}{\begin{picture}(45,30)(0,0)
\SetWidth{1.0}\SetScale{1.0} #1 \end{picture}}\;}
\newcommand{\piccb}[1]{\;\parbox[c]{75pt}{\begin{picture}(75,30)(0,0)
\SetWidth{1.0}\SetScale{1.0} #1 \end{picture}}\;}

\def\Lwidth{1}

\def\Agl(#1,#2)(#3,#4,#5){\PhotonArc(#1,#2)(#3,#4,#5){\Lwidth}
{6.283 #3 mul 360 div #4 #5 sub #4 #5 sub mul sqrt mul Ldensity mul}}
\def\Lgl(#1,#2)(#3,#4){\Photon(#1,#2)(#3,#4){\Lwidth}
{#1 #3 sub #1 #3 sub mul #2 #4 sub #2 #4 sub mul add sqrt Ldensity mul}}
\def\Agh(#1,#2)(#3,#4,#5){\DashArrowArc(#1,#2)(#3,#4,#5){1}}
\def\Aagh(#1,#2)(#3,#4,#5){\DashArrowArcn(#1,#2)(#3,#5,#4){1}}
\def\Lgh(#1,#2)(#3,#4){\DashArrowLine(#1,#2)(#3,#4){1}}
\def\Lagh(#1,#2)(#3,#4){\DashArrowLine(#3,#4)(#1,#2){1}}
\def\Ahh(#1,#2)(#3,#4,#5){\DashCArc(#1,#2)(#3,#4,#5){1}}
\def\Lhh(#1,#2)(#3,#4){\DashLine(#1,#2)(#3,#4){1}}
\def\Aqu(#1,#2)(#3,#4,#5){\ArrowArc(#1,#2)(#3,#4,#5)}
\def\Aaqu(#1,#2)(#3,#4,#5){\ArrowArcn(#1,#2)(#3,#5,#4)}
\def\Lqu(#1,#2)(#3,#4){\ArrowLine(#1,#2)(#3,#4)}
\def\Laqu(#1,#2)(#3,#4){\ArrowLine(#3,#4)(#1,#2)}
\def\Aqq(#1,#2)(#3,#4,#5){\CArc(#1,#2)(#3,#4,#5)}
\def\Lqq(#1,#2)(#3,#4){\ArrowLine(#1,#2)(#3,#4)}
\def\Asc(#1,#2)(#3,#4,#5){\ArrowArc(#1,#2)(#3,#4,#5)}
\def\Lsc(#1,#2)(#3,#4){\ArrowLine(#1,#2)(#3,#4)}
\def\DAsc(#1,#2)(#3,#4,#5){\DashCArc(#1,#2)(#3,#4,#5){3}}
\def\DLsc(#1,#2)(#3,#4){\DashLine(#1,#2)(#3,#4){3}}
\def\TAsc(#1,#2)(#3,#4,#5){\SetWidth{2.0}\CArc(#1,#2)(#3,#4,#5)\SetWidth{1.0}}
\def\TLsc(#1,#2)(#3,#4){\SetWidth{2.0}\ArrowLine(#1,#2)(#3,#4)\SetWidth{1.0}}

\begin{document}

\preprint{ECT*-06-04, HIP-2006-18/TH, TUW-06-02}

\title{The pressure of deconfined QCD
for all temperatures and\\ quark chemical potentials}

\author{A. Ipp}
\affiliation{ECT*, Villa Tambosi, Strada delle Tabarelle 286,\\
I-38050 Villazzano Trento, Italy}
\author{K. Kajantie}
\affiliation{Department of Physics, P.O. Box 64, FI-00014 University of Helsinki, Finland}
\author{A. Rebhan}
\affiliation{Institut f\"ur Theoretische Physik, Technische
Universit\"at Wien, \\Wiedner Hauptstr.~8-10,
A-1040 Vienna, Austria }
\author{A. Vuorinen}
\affiliation{Department of Physics, University of Washington, Seattle, WA 98195, U.S.A.}


\begin{abstract}
  We present a new method for the evaluation of the perturbative
  expansion of the QCD pressure which is
  valid at all values of the temperature and quark chemical potentials
  in the deconfined phase and which we work out up to and including
  order $g^4$ accuracy. Our calculation is manifestly four-dimensional
  and purely diagrammatic --- and thus independent of any effective
  theory descriptions of high temperature or high density QCD.
  In various limits, we recover the
  known results of dimensional reduction and the HDL and HTL
  resummation schemes, as well as the equation of state of
  zero-temperature quark matter, thereby verifying their respective
  validity. To demonstrate the overlap of the various regimes, we
  furthermore show how the predictions of dimensional reduction and
  HDL resummed perturbation theory agree in the regime $T\sim
  \sqrt{g}\mu$. At parametrically smaller temperatures $T\sim g\mu$,
  we find that the dimensional reduction result agrees well with those
  of the nonstatic resummations down to the remarkably low value
  $T\approx 0.2 m_\rmi{D}$, where $m_\rmi{D}$ is the Debye mass at $T=0$. Beyond
  this, we see that only the latter methods connect smoothly to the
  $T=0$ result of Freedman and McLerran, to which the leading
  small-$T$ corrections are given by the so-called non-Fermi-liquid terms,
  first obtained through HDL resummations. Finally, we outline the
  extension of our method to the next order, where it would include
  terms for the low-temperature entropy and specific heats
  that are unknown at present.
\end{abstract}
\maketitle


\tableofcontents

\section{Introduction}

The most fundamental thermodynamic quantity in the theory of strong
interactions, the QCD pressure $p_\rmi{QCD}(T,\mu)$, can at large
values of the temperature $T$ or the quark chemical potentials $\mu$
be computed in a weak coupling expansion in the gauge coupling
constant $g$, defined in the $\tinyMSbar$ renormalization scheme. In
the region where $T$ is larger than all other relevant mass scales in
the problem, the expansion has been extended to include terms of order
$g^6\log g$ \cite{es}-\cite{avpres}, while at $T=0$ and $\mu$ much
greater than the critical chemical potential $\mu_c$, the pressure is
known up to and including terms of order $g^4$ \cite{fmcl}. In between
these regimes, at $0<T\lesssim g\mu$, anomalous contributions from
non-Fermi-liquid behavior have been obtained \cite{Ipp:2003cj},
which involve fractional powers and logarithms of $g$.
The purpose of this paper is to connect all these disjoint
computations through one expression which gives the pressure at all
values of $T$ and $\mu$ up to and including terms of order $g^4$.

Ultimately, the reason for the existing results being valid only in
the separate domains described above is that different computational
methods are practical in different regions of the $\mu$-$T$ plane.
When $T$ is larger than all other dynamical scales, one can, in the
spirit of effective theories, integrate out the degrees of freedom
corresponding to non-zero Matsubara modes (having thermal masses of
order $\pi T$ or higher) to obtain a simpler three-dimensional
effective theory. Formally, the requirement for this is that one must
have\footnote{Due to the chemical potential $\mu$ appearing in the
  fermionic propagator with an imaginary unit, this parameter
  contributes to the long-distance behavior of the free fermionic
  correlation function (\textit{i.e.~}the Fourier transform of the
  free propagator) only through an oscillatory phase factor.
  Therefore, from the point of view of IR physics it can be identified
  as an irrelevant parameter, whose value only affects the validity of
  dimensional reduction through the Debye mass scale. Note, however,
  that this reasoning does not apply to imaginary values of the
  chemical potential, in particular when its magnitude becomes comparable to
  $\pi T$. (In this region one is in any case restricted to values
  $\mu <\pi T /N_c$ due to the loss of periodicity in the pressure as
  a function of imaginary $\mu$ in standard perturbative calculations \cite{rw}. This
  is due to the explicit breaking of the Z($N_c$) symmetry, whose effects are
  otherwise partially visible even in the presence of dynamical fermions.)} $T\gg m_\rmi{D}$
which explicitly excludes the region of $T\lesssim g\mu$. At these
parametrically lower temperatures, perturbation theory requires a different
reorganization which can be most efficiently performed via the hard
dense loop (HDL) approximation. This approach corresponds to a
different effective field theory which is intrinsically
four-dimensional and whose non-local effective action, known only to
one-loop order, was found by Braaten, Pisarski, Taylor and Wong
\cite{BP}. It involves a resummation of nonstatic self-energies which,
in fact, was a crucial ingredient already in the classic calculation
of Freedman and McLerran \cite{fmcl} for the ${\mathcal O}(g^4)$
zero-temperature pressure. In practice, even before reaching this
limit, at $T\sim \mu g^{-5}\exp[-3\pi^2/(\sqrt2 g)]$ \cite{Son:1998uk}, one
of course encounters the non-perturbative pairing instability (color
superconductivity), but in a strictly perturbative calculation these
effects can never be seen.

In view of the fragmented status of the various perturbative results
on the $\mu$-$T$ plane, there is obvious motivation for attempting to develop
an independent and uniform method of calculation which would have the power
of both verifying the validity of all the existing computations and
providing a smooth interpolation between them. Important steps in this direction
have already been taken through recent advances in the analytic calculation of
sum-integrals at arbitrary temperatures and densities
\cite{avpres,antti} as well as in the numerical evaluation of
multi-dimensional integrals involving the one-loop gluon polarization tensor
\cite{ippreb}. In fact, these techniques provide all the required machinery for the
first purely diagrammatic four-dimensional
determination of the QCD pressure up to order $g^4$ at arbitrary $\mu$
and $T$ which is what we set out to perform in this work. We take as
our starting point the systematic analysis of all relevant classes of
Feynman graphs contributing to the partition function up to and
including this order which implies that our calculation
will be independent of any effective descriptions of the fundamental
theory that are valid only in limited regions of the $\mu$-$T$ plane.

As is usual in a weak-coupling analysis, we regard the coupling $g$
as an in principle arbitrarily small parameter, in which we set out to expand
the pressure around the value $g=0$. In practice, the QCD
coupling scale is of course about 1 at 100 GeV and at 1 GeV even
about 2, so for our results to be applicable we must
assume to be working at sufficiently high temperatures and/or densities.
A critical region where the nature of the perturbative
expansion of the pressure changes both qualitatively and quantitatively turns out to
be that of parametrically small $T \lesssim m_\rmi{D} \sim g\mu$, so we have found
it useful to quantify the smallness of $T$ by expressing it in the form $T\sim g^x\mu$.
We then study the changes that occur when $x$ is increased from zero to values larger
than 1, where the pressure can be re-expanded in a series of (fractional)
powers and logarithms of $g$.

Our main results, displayed in Sec.~\ref{sec:numres}, include the
following. In the region $T/\mu \sim g^0$, we explicitly recover the
perturbative expansion of the pressure to order $g^4$ which was derived
in Ref.~\cite{avpres} using dimensional reduction (DR). We
show that this result is valid up to arbitrarily high values of the ratio
$\mu/T$, verifying the non-trivial behavior of the different
perturbative coefficients as functions of the parameter. In deriving these
results, we notice that the larger we choose $\mu/T$ to be, the smaller
values of $g$ we have to use in the series expansion of our numerical result in order
to find agreement with the dimensional reduction one. We interpret this as a
demonstration of the fact that
dimensional reduction amounts to expanding the pressure in powers of $m_\rmi{D}/T$
on top of $g$ and is thus inapplicable in the region $T\lesssim g\mu$.
For temperatures $T\sim \sqrt{g}\mu$, which still fall into the domain of
validity of dimensional reduction, we demonstrate the overlap of the
different existing results by deriving the same expansion for the
pressure (in half-integer powers of $g$) from both the
perturbative results of Ref.~\cite{avpres} and the HDL resummation
scheme of Refs.~\cite{Ipp:2003cj,Gerhold:2004tb}. At parametrically even smaller
temperatures, $T\lesssim g\mu$, we then show how our new
calculation --- the first one entirely independent of the original one
of Freedman and McLerran \cite{fmcl} --- smoothly approaches their
famous result for the $T=0$ pressure, at the same time reproducing the
``anomalous specific heat'' contributions to the pressure obtained previously through
HDL resummations. In contrast, the dimensional reduction result
is observed to agree numerically
remarkably well with those of the nonstatic resummations down to $T\approx 0.2 m_\rmi{D}$, or $2\pi T\approx 1.3m_\rmi{D}$, but
it diverges logarithmically in the limit $T\rightarrow 0$. We explain how this
divergence is related to the behavior of the physical, resummed
expression for the pressure at large but order $g^0$ values of $T/(g\mu)$.

The paper is organized as follows. In Section \ref{sec:prevmeth}, we
review the most important methods applied previously to the evaluation
of the QCD pressure, namely dimensional reduction, HTL/HDL resummations
and zero-temperature perturbation theory.  After this, we present
our new scheme in Section \ref{sec:newappr}, where we identify the
classes of diagrams we need to compute and then write down our new
expression for the pressure as the sum of an analytic power
series in $g^2$ and a logarithmic sum-integral that is to be treated
numerically. In Section
\ref{sec:numres}, we then finally display and discuss our results. We plot
them in various regions of the $\mu$-$T$ plane as characterized by the
exponent $x$ in $T\sim g^x\mu$ and compare our predictions to the
existing results of dimensional reduction \cite{avpres} and HTL/HDL
resummations \cite{Ipp:2003cj,Gerhold:2004tb}. After
this, we summarize our findings in Section \ref{sec:summary}, where we
explain in a detailed fashion the structure of the perturbative
expansion of the QCD pressure on the $\mu$-$T$ plane, and then draw
final conclusions and briefly look into new directions in Section
\ref{sec:concl}.  Several computational details as well as the
explicit analytic forms of many individual contributions to our result are finally
displayed in the Appendices where, in particular, Appendix
\ref{app:num} contains details on the numerical evaluation of the
logarithmic sum-integral introduced in Sec.~\ref{sec:newappr}.

\subsection{The notation}\label{sec:notation}

To conclude the introduction, let us fix our notational conventions which to a large extent follow those of Ref.~\cite{avpres}. First, throughout the text we will be
working in the Feynman gauge, in which many of the existing results our calculation relies on have been derived. The chemical potentials of the different flavors, for which we
apply the notation
\be
\mubar\equiv  \fr{\mu}{2\pi T}, \quad
z \equiv 1/2-\imathb\mubar,
\ee
are assumed to be equal, and thus we suppress all subscripts for them.
(The renormalization scale will be denoted by the symbol $\bar{\Lambda}$
and should not be confused with the above $\mubar$.)
For the analytic result of Eq.~(\ref{panl}), the generalization to several independent chemical potentials is trivial
and only requires the multiplication of the results by $1/N_f \sum_f$, but for the outcome of the numerical calculation, this is less straightforward. All of our methods
can naturally be applied to any combination of $T$ and the $\mu_f$'s, but this will have to be done on a case-by-case basis.

As usual, we denote the most common group theory constants by
\ba
C_A \delta^{cd} & \equiv & f^{abc}f^{abd} \;\,\;=\;\,\; N \delta^{cd}, \\
C_F \delta_{ij} & \equiv & (T^a T^a)_{ij} \;\,\;=\;\,\; \fr{N^2-1}{2N}\delta_{ij},\\
T_F \delta^{ab} & \equiv & \tr T^a T^b \;\,\;=\;\,\; \fr{n_f}{2} \delta^{ab},\\
d_A &\equiv& \delta^{aa} \;\;\,=\;\;\,  N^2 - 1, \\
d_F &\equiv& \delta_{ii} \;\;\,=\;\;\, d_A T_F/C_F \;\;\,=\;\;\, N n_f,
\ea
where the $T_a$ stand for the canonically normalized generators of SU($N$) and the trace is taken over both color and flavor indices.

Our notation for the special functions that appear in the $\e$-expansions of
sum-integrals at finite density follows that of Ref.~\cite{avpres},
which introduced
\ba
\zeta'(x,y) &\equiv& \partial_x \zeta(x,y), \label{specf1}\\
\aleph(n,z) &\equiv& \zeta'(-n,z)+\(-1\)^{n+1}\zeta'(-n,z^{*}), \\
\aleph(z) &\equiv& \Psi(z)+\Psi(z^*).\label{specf3} \ea
Here, $\zeta$
stands for the generalized Riemann zeta function and $\Psi$ for the
digamma function \ba \Psi(w)&\equiv&\fr{\Gamma '(w)}{\Gamma(w)}.  \ea

Finally, we will throughout the following chapters apply dimensional regularization at $d=4-2\e$ dimensions and (unless otherwise stated) in Euclidean metric, using as
the momentum integration measure
\ba
\int_p \;\,\equiv\;\, \int\! \fr{{d}^{d-1} p}{(2\pi)^{d-1}} &=& \Lambda^{2\e}\!\!\int\!
\fr{{d}^{d-1} p}{(2\pi)^{d-1}}\;\,\equiv\;\, \(\!\fr{e^{\gamma}\bar{\Lambda}^2}{4\pi}\!\!\)^{\!\!\e}\!\!\int\!
\fr{{d}^{d-1} p}{(2\pi)^{d-1}}, \label{asdfg}\\
\sumint_{P/\{P\}} &\equiv& T \sum_{p_0/\{p_0\}} \int_p,
\ea
where $\bar{\Lambda}\equiv (4\pi e^{-\gamma})^{1/2}\Lambda$ is the scale parameter of the
$\msbar$ scheme. The symbol $P/\{P\}$ refers to bosonic and fermionic
momenta, respectively, for which capital letters stand for four-momenta and lower-case ones for
three-momenta. Using these definitions, bosonic and fermionic 1-loop master integrals
are defined by
\be
\label{ints1}
{\cal I}_{n}^m \equiv \sumint_P \fr{\(p_0\)^m}{\(P^2\)^n}, \quad
\widetilde{\cal I}_{n}^m \equiv \sumint_{\{P\}} \fr{\(p_0\)^m}{\(P^2\)^n}.
\ee
The bosonic and fermionic distribution functions at temperature $T$ and chemical potential $\mu$ are denoted by
\begin{eqnarray}
n_{b}(k) & = & \frac{1}{e^{k/T}-1},\label{nb}\\
n_{f}(k) & = & \frac{1}{2}\left(\frac{1}{e^{(k-\mu)/T}+1}+\frac{1}{e^{(k+\mu)/T}+1}\right).\label{nf}
\end{eqnarray}

\section{Previous methods}\label{sec:prevmeth}

Let us begin by reviewing the main results available for the QCD pressure through various previous perturbative calculations in order to summarize the current
understanding of the behavior of the quantity in various regions of the $\mu$-$T$ plane.
The methods to be covered include dimensional reduction which has been used to determine the pressure up to and including order $g^6\ln\,g$
at finite $T$ and $\mu$ \cite{klry,avpres}, the perturbative $T=0$ techniques of Freedman and McLerran \cite{fmcl}, and
the HDL resummations that have proven to be highly effective at temperatures non-zero but parametrically smaller than $\mu$ \cite{Ipp:2003cj,Gerhold:2004tb}.

Recently there has been some progress to include effects
of nonzero quark masses in the dimensional reduction approach
\cite{Laine:2006cp}
as well as to the next-to-leading order pressure at zero temperature
\cite{Fraga:2004gz}.
In the following we shall always restrict our attention to the case of massless
quarks.

\subsection{Dimensional reduction}

In finite-temperature QCD, dimensional reduction is based on the
observation that at sufficiently high temperatures one has the
parametric scale hierarchy $m_\rmi{mag}\ll m_\rmi{D}\ll \pi T$ between
the magnetic mass $\sim g^2 T$, the Debye mass $\sim g T$, and the smallest non-zero Matsubara
frequency, respectively. One may then integrate out all the non-static
degrees of freedom from the original theory (corresponding to the last of
the above energy scales), leaving behind a
three-dimensional effective theory for the static gluon fields. At
leading order, this integration-out can be performed explicitly, but
beyond that it becomes more convenient to start from the most generic
possible three-dimensional Lagrangian that respects the correct
symmetries and to determine the values of its parameters by matching a
set of simple physical quantities to their values in the full theory.
This leads to a Lagrangian for the effective theory, commonly dubbed
electrostatic QCD (EQCD), of the (Euclidean) form \cite{bn1,Hart:2000ha}
\ba {
  \mathcal
  L}_{\rmi{EQCD}}&=&\fr{1}{2}\tr F_{ij}^2+\tr\!\big[(D_iA_0)^2\big]+m_\rmi{E}^2\tr
A_0^2\nn
&+&\fr{ig^3}{3\pi^2}\sum_f \mu_f\tr A_0^3
+\lambda_{\rmi E}^{(1)}\(\tr A_0^2\)^2 +\lambda_{\rmi E}^{(2)}\tr A_0^4
+\delta{\mathcal L}_\rmi{E}.
\label{LEQCD} \ea

In the above Lagrangian, $m_{\rmi E}$
agrees to leading order with the physical Debye mass $m_\rmi{D}$,
\ba
m_\rmi{D}&=&g\sqrt{\fr{C_A+T_F}{3}T^2+T_F\fr{\mu^2}{\pi^2}},
\ea
and
the last term $\delta{\mathcal L}_\rmi{E}$ corresponds to
higher-dimensional operators that can be neglected for most practical
purposes, as they contribute \textit{e.g.~}to the
pressure starting only at ${\mathcal O}(g^7)$.
The cubic, $C$-odd term \cite{KorthalsAltes:1999cp,Hart:2000ha}
in Eq.~(\ref{LEQCD}) is only present at finite
chemical potentials and contributes to the pressure at order $g^6 \ln\,g$,
(but turns out to be responsible for the {\em leading} order result of the
off-diagonal quark number susceptibilities at zero chemical
potential \cite{Blaizot:2001vr}). The fields have been rescaled from their four-dimensional
values by $\sqrt{T}$ to have the canonical dimensionality at $d=3$,
and the field strength and covariant derivative contain the
three-dimensional gauge coupling $g_\rmi{E}\equiv g\sqrt{T}+{\mathcal
  O}(g^3)$. The new theory should be able to describe the physics of
the original one at distance scales of order $1/(gT)$ and higher.

Assuming $g$ to be sufficiently small (\textit{i.e.}~$T$ sufficiently high), the above integrating-out procedure can be continued one step further by removing also the massive longitudinal
gluon $A_0$ from the theory, thus producing an effective three-dimensional pure Yang-Mills theory for the $A_i$ fields. This theory, magnetostatic QCD (MQCD), is defined by the Lagrangian
\ba
{\cal L}_\rmi{MQCD} & = & \fr12 \tr F_{ij}^2,
\ea
with the gauge coupling constant $g_\rmi{M}=g_\rmi{E}+{\mathcal O}(g_\rmi{E}^3/m_\rmi{E})$ appearing in
the field strength tensor. The perturbative expansion of the pressure of the full theory
can then be given in the form
\ba
p&=&p_\rmi{E}+p_\rmi{M}+p_\rmi{G},
\ea
where the first term corresponds to the coefficient of the unit operator of EQCD, not explicitly written in Eq.~(\ref{LEQCD}), and can be evaluated through a strict perturbative
expansion (implying no resummations) of the full theory pressure. The second term, $p_\rmi{M}$, represents the same thing for MQCD and can be computed through a
diagrammatic expansion of the pressure of this theory, while the last piece, $p_\rmi{G}$, corresponds to the (only) fundamentally non-perturbative contribution to the
full theory pressure in
the form of the pressure of MQCD. To order $g^6\ln\,g$ --- the current state-of-the-art of the field --- the three terms have been computed at
$\mu=0$ in Ref.~\cite{klry} and at $\mu\neq 0$ in Ref.~\cite{avpres}.
In Eq.~(\ref{dimredpr5}) of Appendix A we quote the result of Ref.~\cite{avpres} to order
$g^5$ in a simplified case where equal chemical potentials have been assumed for all
quark flavors.

Formally, the condition ensuring the validity of dimensional reduction is that the temperature is the largest fundamental energy scale in the problem, so that the above scale hierarchies
are satisfied. In the limit of low
temperatures $T\ll\mu$, the Debye mass becomes just $m_\rmi{D}=\sqrt{N_f/2}g\mu/\pi$ which shows that the identification of the static degrees of freedom as the only IR sensitive ones is
justified as long as $T\ll g\mu$ or, formally, $T\sim g^x \mu$, with any $x<1$. Assuming this to be the case and defining $\tau\equiv\pi T/(g^x \mu )$, one can then extend the validity
of the standard dimensional reduction results to the case of an arbitrary $x<1$ by simply inserting this value of the temperature,
\ba\label{tau}
\pi T&=&\tau g^x\mu,
\ea
into Eq.~(3.13) of Ref.~\cite{avpres}
and expanding the result as a power series in $g$ up to but not including order $g^6\mu^4$. The expansions of the $\aleph$ functions appearing
in the coefficient of the $g^4$ term with an argument $\mu/(2\pi T)$ can be performed with the aid of the results of the Appendix D of the same reference.

Here, we exhibit the result of the above procedure only for the case of $x = 1/2$, and as above we
assume all quark flavors to have the same chemical potential\footnote{Here, one cannot simply multiply the result by $\fr{1}{N_f}\sum_f$ in order to generalize it to an arbitrary
number of flavors with independent chemical potentials.} $\mu$,
where we can use the simplified result of Eq.~(\ref{dimredpr5}).
After some straightforward manipulations we obtain
\ba\label{pDRsqrtg}
p_\rmi{DR}&=&
\fr{d_A\mu^4}{(4\pi)^2}\Bigg\{\fr{4d_F}{3d_A}+g\fr{8\tau^2d_F}{3d_A}-g^2\bigg(\fr{T_F}{2\pi^2}-\fr{16\tau^4}{45}\Big(1+\fr{7d_F}{4d_A}\Big)\bigg)\nn
&-&g^3\fr{T_F\tau^2}{\pi^2}+g^{7/2}\fr{4T_F^{3/2}\tau}{3\pi^3} \nn
&-&\fr{g^4}{576\pi^4}\bigg(72T_F^2\,\ln\,g+32(2C_A+5T_F)\pi^2\tau^4+4C_AT_F\Big(71-33\,\ln\,2+33\,\ln\,\fr{\bar{\Lambda}}{\mu}\Big)\nn
&-&153C_FT_F+4T_F^2\Big(11-36\gamma+36\,\ln\,8\tau-12\,\ln\,\fr{\bar{\Lambda}}{\mu}\Big)\bigg)\nn
&+&g^{9/2}\fr{2\sqrt{T_F}(C_A+T_F)\tau^3}{3\pi^2}\nn
&+&\fr{g^5\tau^2T_F}{288\pi^4}\bigg(4(53C_A-6C_F+2T_F) \ln\,g
+3C_F\Big(35+16\,\fr{\zeta '(-1)}{\zeta (-1)}-16\,\ln\,2\tau\Big)\nn
&+&C_A\Big(-227-20\,\ln\,2+48\gamma+104\fr{\zeta '(-1)}{\zeta
(-1)}+144\,\ln\,\fr{T_F}{\pi^2}-152\,\ln\,\tau-132\,\ln\,\fr{\bar{\Lambda}}{\mu}\Big)\nn
&-&4T_F\Big(5+8\,\ln\,2-12\gamma+16\fr{\zeta '(-1)}{\zeta(-1)}-12\gamma-4\,\ln\,\tau-12\,\ln\,\fr{\bar{\Lambda}}{\mu}\Big)\bigg)\nn
&+&\fr{g^{11/2}\tau}{72\sqrt{T_F}\pi^5}\bigg(-33C_AT_F^2\,\ln\,g+4C_A^2\pi^2\tau^4+8C_AT_F\pi^2\tau^4+4T_F^2\pi^2\tau^4-54C_FT_F^2\nn
&+&33C_AT_F^2\Big(1+2\gamma-2\,\ln\,4\tau+2\,\ln\,\fr{\bar{\Lambda}}{\mu}\Big)
+12T_F^3\Big(1+2\,\ln\,2-2\,\ln\,\fr{\bar{\Lambda}}{\mu}\Big)\bigg)\nn
&+&\fr{g^6\,\ln\,g\,\tau^4}{864\pi^4}\bigg(166C_A^2+\fr{832}{5}C_AT_F+72C_F
T_F+\fr{371}{5}T_F^2\bigg)
+O(g^6)
\Bigg\}.
\ea
To test the validity of this expression, one may compare it with the prediction of other approaches entirely independent of dimensional reduction which we will do in Sec.~\ref{sec:numres}.

\subsection{The Freedman-McLerran result for the $T=0$ pressure}

At exactly zero temperature, important simplifications take place which resulted in the perturbative expansion of the QCD pressure being extended
to order $g^4$ in this region already remarkably early, in the late 1970's \cite{fmcl}. The most important effects stem from the fact that at $T=0$ loop integrations
become much more straightforward than at finite temperature due to the Matsubara sums reducing to ordinary integrals.
The computation of Ref.~\cite{fmcl} was organized in much the same way as the one to be presented in this paper, with the exception that most of the technical details were much
less involved. Especially with zero quark masses, the evaluation of the 2PI graphs at zero temperature is significantly simpler than at $T\neq 0$, leaving as the real
challenge the computation of the 'plasmon' sum, i.e.~the resummation of ring diagrams. This they did in a highly imaginative --- and to a large extent
analytic --- way, having to resort to numerics only in evaluating a few low-dimensional integrals.

While the original calculation of Ref.~\cite{fmcl} was performed in the momentum subtraction scheme and moreover
using by now somewhat out-dated methods that led to largish
numerical error bars, the result was to a large extent independently\footnote{With the exception of the plasmon sum, for which similar methods as those in
Ref.~\cite{fmcl} were used. A translation of the original result
of Ref.~\cite{fmcl} to the $\tinyMSbar$ scheme was previously
given in the last reference of Ref.~\cite{BIR} and in Ref.~\cite{Fraga:2001id}.}
rederived in Ref.~\cite{avpres} in the $\tinyMSbar$ scheme. For $N_f$ flavors of quarks at a equal chemical potential $\mu$, the result reads
\ba\label{pFMcL}
p(\mu,T=0)&=&\fr{d_A\mu^4}{2\pi^2}\Bigg(\fr{d_F}{6d_A}-\fr{g^2}{(4\pi)^2}T_F-\fr{g^4}{(4\pi)^4}T_F\bigg[4T_F\,\ln\,\fr{g^2T_F}{(4\pi)^2}+\fr{2}{3}(11C_A-4T_F)\ln\,\fr{\bar{\Lambda}}{\mu}\nn
&-&\fr{17}{2}C_F+\fr{C_A}{36}(568-264\,\ln\,2)\nn
&-&2T_F\(\fr{88}{9}-14\,\ln\,2+\fr{16}{3}(\ln\,2)^2-\delta-\fr{2\pi^2}{3}\)\bigg]+ O(g^6\ln g)\Bigg) ,
\ea
where $\delta\approx -0.856383209...$ is defined through a one-dimensional integral in Ref.~\cite{avpres}.

\subsection{Minimal HTL/HDL resummation}

At low but non-vanishing temperatures, the Freedman-McLerran
result for the zero-temperature interaction pressure (where the
Stefan-Boltzmann (SB) result has been subtracted off) to order $g^4$
needs to be corrected by $T$-dependent contributions.
At two-loop order there are infrared safe terms
of order $g^2\mu^2 T^2$ which clearly are more important
than the unknown four-loop $T=0$ terms $\sim g^6\mu^4$ as long as $T \gtrsim g^2\mu$.
On the other hand, when $T$ approaches the scale $g\mu$ from the dimensional reduction region above,
the leading contribution of ring
diagrams $T m_\rmi{D}^3\sim g^3 T\mu^3$ becomes comparable
to $g^2\mu^2 T^2$. A naive extrapolation of the $T m_\rmi{D}^3$ term
to temperatures parametrically smaller than $g\mu$ would
even suggest that for $T\sim g^x \mu$ with $x>1$ these become
more important than the two-loop terms, as $T m_\rmi{D}^3 \sim g^{3+x}\mu \gg g^{2+2x}\mu
\sim g^2\mu^2 T^2$. As we shall see, the $T$-dependent
contributions from ring diagrams indeed become more important
than the 2-loop term $g^2\mu^2 T^2$ here, though they are not enhanced
by a relative factor $g^{-(x-1)}$ as suggested by dimensional
reduction, but instead only by a logarithm.

For temperatures $T \lesssim g\mu$, where dimensional reduction
is no longer applicable, it becomes important to keep the nonstatic
parts of the gluon self energy in the ring diagrams.
At low momenta and frequencies of the order $g\mu$, the leading
terms in the gluon self-energy are given by the
so-called hard thermal loops (HTL) approximation with the overall
$m_\rmi{D}^2$ factor replaced by its zero-temperature value ---
a special case occasionally referred to as hard dense loops (HDL).
In the longitudinal gluon propagator, one can observe
the usual Debye screening effect at the frequency $\omega\ll g\mu$,
but in the transverse propagator the situation
is more complicated. At strictly zero frequency the
magnetostatic HDL propagator is massless, but for
small but nonvanishing frequencies
$\omega\ll q\lesssim m_\rmi{D}$ its inverse has the form
\ba
q^2-\omega^2+\Pi^\rmi{HDL}_\rmi{T}(\omega,q)
&=&q^2-{i\pi m_\rmi{D}^2\04}{\omega\0q}+O(\omega^2).
\ea
The transverse part of the propagator thus has a pole
at imaginary $q$ and $|q|=m_\rmi{mag}(\omega)$,
introducing a new parametrically small dynamical
screening mass \cite{Weldon:1982aq,Kraemmer:2003gd}
\ba
\label{mmdyn}
m_\rmi{mag}(\omega)&=&\left( {\pi m_\rmi{D}^2 \omega \0 4}
\right)^{1/3},\qquad \omega\ll m_\rmi{D},
\ea
which represents an in-medium version of Lenz's law.
As soon as the temperature is small but nonvanishing, the ring
diagrams obtain contributions involving the Bose-Einstein
distribution function which leads to sensitivity
to this additional scale.
In these contributions, we effectively have
$m_\rmi{mag}(\omega\sim T) \sim g^{(2+x)/3}\mu$
for $T\sim g^x\mu$ and $x>1$.
Note that
this is parametrically smaller than $m_\rmi{D}\sim g\mu$, but always larger than
the magnetic mass scale of MQCD, $m_\rmi{mag}(\omega\!=\!0)=
g^2 T \sim g^{2+x}\mu$.

The resummation of the nonstatic transverse gluon self-energy
gives rise to terms nonanalytic in the temperature which to lowest
order in a low-temperature expansion turn out to be of the order
$g^2 \mu^2 T^2 \ln\,T$. This gives rise to
so-called anomalous or non-Fermi-liquid behavior in the
entropy and specific heat at low $T$, because instead of
the usual linear behavior in $T$ the entropy then has a $T\ln\,T$
term which is the hallmark of a breakdown of the Fermi-liquid
picture (first discussed in the context
of nonrelativistic QED by Norton, Holstein and Pincus \cite{Holstein:1973}).
Indeed, inspection of the dispersion laws of
fermionic quasiparticles reveals that there is a logarithmic
singularity in the group velocity at the therefore no longer
sharply defined Fermi surface\footnote{A systematic
calculation of the group velocity beyond the leading-log approximation has only
recently been carried out in Ref.~\cite{Gerhold:2005uu}.}.

For a long time, only the multiplicative coefficient of the $T\ln\,T$ term
in the specific heat was known. It was only rather recently \cite{Ipp:2003cj}
that also the scale under the logarithm was determined
together with the next order terms in the low-temperature ($T\ll g\mu$)
series which in addition involves fractional
powers of $T$ due to the cubic root in Eq.~(\ref{mmdyn}).
For the pressure, these ``anomalous'' $T$-dependent contributions
are contained in an expression,
which was first derived in Ref.~\cite{Gerhold:2004tb}
and which we shall label by HDL$^+$,
\ba\label{PHDL}
{1\0N_g}\delta p^\rmi{HDL$^+$}
&=&-{g^2T_F\048\pi^2}\mu^2T^2
-{1\02\pi^3}\int_0^\infty dq_0\, n_b(q_0)
\int_0^\infty dq\,q^2\,\biggl[ 2\,\im \ln \left( q^{2}-q_{0}^{2}+\Pi^\rmi{HDL} _\rmi{T} \0 q^{2}-q_{0}^{2} \right )\nonumber\\
&+&\im \ln \left( \frac{q^{2}-q_{0}^{2}+\Pi^\rmi{HDL} _\rmi{L}}{q^{2}-q_{0}^{2}}\right)
\biggr]  + O(g^2T^4) + O(g^3 \mu T^3)+ O(g^4\mu^2T^2),
\ea
with $\delta p$ denoting the temperature-dependent part of the interaction
pressure
\ba\label{deltaDeltap}
\delta p & \equiv & \Delta p - \Delta p|_{T=0}, \nn
\Delta p & \equiv & p - p_\rmi{SB}.
\ea
The expression (\ref{PHDL}) can be viewed as a minimal\footnote{As
opposed to the HTL/HDL resummation considered in \cite{ABS,BIR}
which aims at improving the convergence of the perturbative
series at high temperature
by retaining higher-order effects from HTL/HDL physics
beyond what is needed from a perturbative point of
view. The + in HDL$^+$ and HTL$^+$
is meant as a reminder that the
corresponding quantities are not expressed in terms
of HTL/HDL quantities only, but combined with
unresummed infrared-safe contributions.}
resummation of HDL diagrams, where the HDL self-energies are only kept in
the infrared sensitive part of the ring diagrams involving the distribution function $n_b$, while
infrared safe two-loop contributions are treated in an unresummed form.

\subsubsection{$T$ parametrically smaller than $m_\rmi{D}$}

With $g\ll 1$ and $x>1$ in $T\sim g^x\mu$, the temperature is parametrically
smaller than the Debye mass $m_\rmi{D}\sim g\mu$ and
Eq.~(\ref{PHDL}) contains the leading contributions
to the temperature-dependent parts of the interaction pressure,
which ignoring logarithms are of order $g^2\mu^2 T^2\sim g^{2+2x}\mu^4$,
while the higher-order terms in Eq.~(\ref{PHDL}) are at least of
order $g^{4+2x}\mu^4$.
The Freedman-McLerran result for the $T=0$ pressure, Eq.~(\ref{pFMcL}),
is accurate to order $g^4\mu^2$ and its error is of order $g^6\mu^4$
(again ignoring logarithms of $g$).
Eq.~(\ref{PHDL}) thus represents the leading correction to the
Freedman-McLerran result
as long as $x<2$ (i.e., $T\gtrsim g^2\mu$), whereas
in quantities such as the entropy density $s=\6p/\6T$ and the various
specific heats, where the $T=0$ part of the pressure drops out,
it is in fact the leading term in the interaction part for all $x\ge1$.

\def\g{\bar g } 
\def\t{\bar\tau } 
In Eq.~(\ref{PHDL}), $g$ appears only in the
combination $\g^2\equiv g^2 T_F$, and it is therefore convenient
to define a reduced temperature variable
\be\label{bartau}
\t=\pi T/(\g^x \mu).
\ee
For $x>1$, the perturbative content of Eq.~(\ref{PHDL}) is that
given by the low-temperature expansion worked out in
Refs.~\cite{Ipp:2003cj,Gerhold:2004tb}. With the above variables, this reads
\ba\label{PHDLx}
{1\0N_g}{\delta p^\rmi{HDL$^+$}\0m_\rmi{D}^4}
&=&  {\t^2\g^{2(x-1)}\over72}\left(\ln\left({1\0\t \g^{x-1}}\right)
+\ln{4\0\pi}
  +\gamma_E-{6\over\pi^2}\zeta^\prime(2)-{3\02}\right)\nonumber\\
  &-&{2^{2/3}\Gamma\left({8\over3}\right)\zeta\left({8\over3}\right)\over3\sqrt{3}\pi^{7/3}}
  \t^{8/3}\g^{8(x-1)/3}
  +8{2^{1/3}\Gamma\left({10\over3}\right)\zeta\left({10\over3}\right)
  \over9\sqrt{3}\pi^{11/3}}\t^{10/3}\g^{10(x-1)/3}\nonumber\\
  &+&{2048-256\pi^2-36\pi^4+3\pi^6\over2160\pi^2}\t^4 \g^{4(x-1)}
  \left[\ln\left({1\0\t\g^{x-1}}\right)+\ln\pi+\bar c\, \right]
\nonumber\\
&+&{O}(\g^{14(x-1)/3})
+{O}(g^{2x}),
\ea
where $\bar c\approx 4.0993485
\ldots$ is given by a numerical integral
defined in Ref.~\cite{Gerhold:2004tb}.
The latter of the error terms in Eq.~(\ref{PHDLx}) corresponds to the
leading-order terms to be expected from three- and higher-loop contributions\footnote{There,
$g^2$ no longer appears exclusively in combination with $T_F$.}
proportional to $g^4\mu^2 T^2$ which are presumably
enhanced by logarithms of $T$ and $g$. Depending on the value of $x>1$, a finite number of terms in the
low-$T$ expansion remain more important than this
(see Fig.~\ref{fig:orders} in Sec.~\ref{sec:summary}).

When $x=1$, i.e.\ $T\sim g\mu$, the expansion of Eq.~(\ref{PHDLx}) clearly breaks
down (unless $\t\ll 1$) and the HDL-resummed expression of Eq.~(\ref{PHDL})
therefore needs to be evaluated numerically as in Ref.~\cite{Gerhold:2004tb}.
This expression has then the form of $g^4\mu^4$ times
a function of $T/(g\mu)$, and is therefore of the same order as
the $g^4$ term of the $T=0$ pressure of Freedman and McLerran, to
which it is to be added. As displayed in Ref.~\cite{Gerhold:2004tb}
for the case of the entropy,
and as we shall see for the pressure in the plots of Section \ref{sec:numres}
of the present paper,
the $T$-dependent terms of Eq.~(\ref{PHDLx}) smoothly
interpolate between a dominant $g^2 T^2\mu^2\ln\,T$ behavior at low temperature
and the terms of order $g^2T^2 \mu^2$, $g^3 \mu^3 T$, and $g^4\mu^4\ln\,T$
of the dimensional reduction pressure which
should be the dominant terms at sufficiently high temperatures and which
remain comparable to $g^4\mu^4$ as long as the parametric equality $T\sim g\mu$ holds.

\subsubsection{$T$ parametrically larger than $m_\rmi{D}$}

When $x<1$ in $T\sim g^x\mu$, \textit{i.e.}~$T\gg g\mu$,
dimensional reduction provides the most accurate
result available for the QCD pressure. Up to an error of the order of
three-loop contributions proportional to $g^4\mu^2 T^2\sim g^{4+2x}\mu^4$,
one can however reproduce its prediction
by extending the above HDL-resummed calculation
to include the leading thermal corrections to the gluon self-energy. In practice,
this means replacing the HDL approximation by the HTL one and also keeping the order $g^2 T^4$
terms originating from infrared-safe two-loop contributions to the pressure that were
omitted in Eq.~(\ref{PHDL}) because they were of too high order when $x\ge1$.
This possibility was mentioned in Ref.~\cite{Gerhold:2004tb}, but
not considered further because that work concentrated
on the region of $T\lesssim g\mu$. For the purposes of the present paper, we however
write down the straightforward extension of Eq.~(\ref{PHDL}) to the HTL approximation
in the form
\ba\label{PHTL}
{1\0N_g}\delta p^\rmi{HDL$^+$}&=&-{g^2T_F\048\pi^2}\mu^2T^2+{g^2(2C_A-T_F)\0288}T^4\nn
&&-{1\02\pi^3}\int_0^\infty dq_0\, n_b(q_0)
\int_0^\infty dq\,q^2\,\biggl[ 2\,\im \ln \left( q^{2}-q_{0}^{2}+\Pi^\rmi{HTL} _\rmi{T} \0 q^{2}-q_{0}^{2} \right )\nonumber\\&&\qquad
+\im \ln \left( \frac{q^{2}-q_{0}^{2}+\Pi^\rmi{HTL} _\rmi{L}}{q^{2}-q_{0}^{2}}\right)
\biggr] + O(g^4\mu^2T^2).
\ea

Combining the above expression with the Freedman-McLerran result of Eq.~(\ref{pFMcL}) to obtain
\ba\label{PHTLtot}
\Delta p^\rmi{HDL$^+$} &\equiv & p^\rmi{HDL$^+$} - p_\rmi{SB} \;\;\equiv\;\;
\Delta p^\rmi{FMcL}+\delta p^\rmi{HDL$^+$},
\ea
we have an expression for the interaction pressure
whose error is of order $g^{{\rm min}(4+2x,6)}$
for all $T\sim g^x \mu$. This we shall compare (and thus test) in the following
with our new approach which resums the complete one-loop gluon self-energy
(i.e., not only the leading HTL/HDL contribution) in ring diagrams.
Note that the accuracy of (\ref{PHTLtot}) is at least of order $g^4$
for all parametrically small temperatures, excluding only the case of $x=0$,
where $T\sim \mu$.

\section{The new approach}\label{sec:newappr}

In this Section, we introduce our novel and strictly four-dimensional
calculational scheme designed to reproduce the perturbative expansion
of the QCD pressure up to and including order $g^4$ at all values of
$\mu$ and $T$.
Our guiding principle is that when faced with the necessity to
sum up graphs with multiple self energy insertions to circumvent
infrared problems, we consider the entire self energy
and not only those parts which are identified as relevant
in some effective field theory description,
such as the Debye mass in dimensional reduction or
the HTL/HDL self energy in the corresponding resummation schemes.
Because we (at present) limit ourselves to order $g^4$ accuracy,
it will be sufficient to resum only one-loop self-energies in the
infrared sensitive graphs, while IR safe diagrams will be treated
perturbatively, using bare propagators.
This will introduce gauge dependence to our results, but only at orders
beyond $g^4$ which we will explicitly discard by either considering values
of $g$ low enough for the higher order terms to be negligible or by performing
numerical series expansions up to ${\mathcal O}(g^4)$.

We begin our treatment with a general diagrammatic analysis where we identify
the relevant classes of Feynman graphs that need to be considered. After that,
we describe their evaluation and show how adding them together leads to the
final result displayed in Section \ref{subsec:result}. Many details of the
calculations as well as the results of several individual pieces of the result
are left to be covered in the Appendices.

\subsection{Identification of the relevant diagrams}

\def\Elmeri(#1,#2,#3){{\pic{#1(15,15)(15,0,180)%
 #2(15,15)(15,180,360)%
 #3(0,15)(30,15)}}}

\def\Petteri(#1,#2,#3,#4,#5,#6){\pic{#3(15,15)(15,-30,90)%
 #1(15,15)(15,90,210)%
 #2(15,15)(15,210,330) #5(2,7.5)(15,15) #6(15,15)(15,30) #4(15,15)(28,7.5)}}

\def\Jalmari(#1,#2,#3,#4,#5,#6){\picc{#1(15,15)(15,90,270)%
 #2(30,15)(15,-90,90) #4(30,30)(15,30) #3(15,0)(30,0) #5(15,0)(15,30)%
 #6(30,30)(30,0) }}

\def\Oskari(#1,#2,#3,#4,#5,#6,#7,#8){\picc{#1(15,15)(15,90,270)%
 #2(30,15)(15,-90,90) #4(30,30)(15,30) #3(15,0)(30,0) #6(15,0)(15,15)%
 #5(15,15)(15,30) #8(30,30)(30,15) #7(30,15)(30,0) }}

\def\Sakari(#1,#2,#3){\picb{#1(15,15)(15,30,150)%
#1(15,15)(15,210,330) #2(0,15)(7.5,-90,90) #2(0,15)(7.5,90,270) %
#3(30,15)(7.5,-90,90) #3(30,15)(7.5,90,270) }}

\def\Maisteri(#1,#2){\picb{#1(15,15)(15,0,150)%
#1(15,15)(15,210,360) #2(0,15)(7.5,-90,90) #2(0,15)(7.5,90,270) #1(37.5,15)(7.5,0,360) }}

\def\Tohtori(#1,#2){\picb{#1(15,15)(15,0,360)#1(45,15)(15,0,360) }}

\def\Pietari(#1){\pic{#1(15,15)(15,0,360) #1(15,1)(20,40,140)#1(15,29)(20,220,320)}}

\def\Ari(#1,#2){\pic{#1(15,15)(15,0,360) #2(0,15)(27,22) #2(0,15)(27,8)}}

\def\Kari(#1){\pic{#1(15,15)(15,-90,270)}}

\begin{figure}[t]
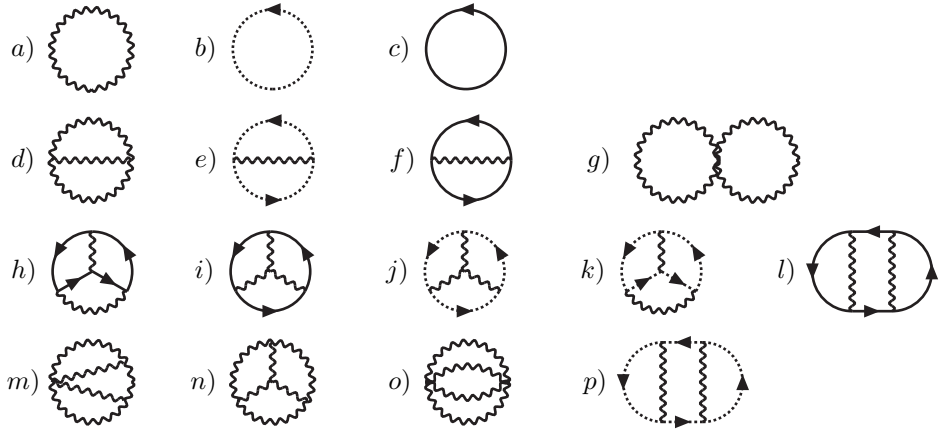

\centering
\ba \nonumber
\begin{array}{lllll}
a)~ \Kari(\Agl) & b)~
\Kari(\Agh)
\;\;\;\; c)~
\Kari(\Asc)
\nn
\nn
d)~ \Elmeri(\Agl,\Agl,\Lgl) & e)~
\Elmeri(\Agh,\Agh,\Lgl)
\;\;\;\; f)~
\Elmeri(\Asc,\Asc,\Lgl)
\;\;\;\; g)~
\!\!\!\!\Tohtori(\Agl,\Agl)
\nn
\nn
h)~\Petteri(\Asc,\Agl,\Asc,\Lsc,\Lsc,\Lgl)
&i)~\Petteri(\Asc,\Asc,\Asc,\Lgl,\Lgl,\Lgl)
\;\;\;\;
j)~
\Petteri(\Agh,\Agh,\Agh,\Lgl,\Lgl,\Lgl)
\;\;\;\; k)~
\Petteri(\Agh,\Agl,\Agh,\Lgh,\Lgh,\Lgl)
\;\;\;\; l)~
\Jalmari(\Asc,\Asc,\Lsc,\Lsc,\Lgl,\Lgl)
\nn
\nn
\!m)~
\!\Ari(\Agl,\Lgl)
&\!n)~
\Petteri(\Agl,\Agl,\Agl,\Lgl,\Lgl,\Lgl)
\;\;\;\;  o)~
\Pietari(\Agl)
\;\;\;\; p)~
\Jalmari(\Agh,\Agh,\Lgh,\Lgh,\Lgl,\Lgl)
\nn
\end{array}
\ea
\caption[a]{The one-, two- and three-loop two-gluon-irreducible (2GI) graphs of QCD. The wavy line stands for a gluon, the
dotted line a ghost and the solid line a quark.}
\end{figure}

To determine the QCD pressure up to and including order $g^4$ on the entire
deconfined phase diagram of the theory, our first task is to identify all
Feynman diagrams that contribute to the partition function at this order.
These trivially include the two-{\em gluon}-irreducible (2GI) diagrams
up to three-loop order, displayed in Fig.~1 which a straightforward power
counting as well as the explicit calculation of
Ref.~\cite{avpres} confirms as infrared finite for all temperatures and
chemical potentials.

In addition to these cases, there are, however, several other classes of IR
sensitive diagrams that need to be resummed to infinite loop order, as a
power counting reveals that the dressing of (at least some of) their gluon lines with an arbitrary
number of one loop gluon polarization tensors does not increase their order beyond $g^4$. These
diagrams are shown in Fig.~2, where the first set corresponds to the well-known class
of ring diagrams that leads to the known $g^3$ and $g^4\ln\,g$ contributions to the
pressure at high $T$ \cite{jk,tt} and to the $g^4\ln\,g$ term at $T=0$ \cite{fmcl}.
Among others, this class contains the set of all three-loop two particle reducible (2PR)
graphs of the theory which are missing from Fig.~1.

As we shall see (in contradiction to the opposite assertion in Ref.~\cite{jk}),
the resummation of the ring diagrams is, however, not enough to obtain the entire
order $g^4$ term correctly at nonzero $T$.
Although without resummation starting at orders $g^6$, $g^8$ and $g^6$, respectively,
the classes of Fig.~2 b-d, corresponding to
self-energy insertions in the gluonic two-loop 2GI diagrams 1d and 1g,
have the potential to give rise to contributions of order
${\mathcal O}(g^4T^2\mu^2)$ and ${\mathcal O}(g^4T^4)$ to the pressure.
When $T$ is not parametrically larger than $m_\rmi{D}$, it turns out that
only the class b gives a non-zero contribution at this order, being proportional to
$g^2T^2m_\rmi{D}^2$. When $T\sim g^x\mu$ with $x>0$, none of the three classes
contributes to the pressure to order $g^4\mu^4$, but in the calculation of the
low-temperature entropy and specific heat they have to be taken into account already at
order $g^4 \mu^2 T$.

For any other classes of diagrams apart from those shown in Figs.~1 and 2, it is very straightforward to
see that the contributions will be beyond order $g^4$. In particular, if we were to add an additional
gluon line with some number of self energy insertions into the graphs of Fig.~2 b-d (\textit{i.e.~}dressing
the three-loop 2GI diagrams with self energies), we would notice that
the two extra insertions of the coupling constant due to the new vertices (vertex) ensure that these graphs only contribute
to the pressure at order $g^6$. Similarly, one can see that the inclusion of the two-loop self energy into the ring
diagrams only has an effect on the pressure starting at ${\mathcal O}(g^5)$.

\subsection{The 2GI diagrams}

In Feynman gauge,
the sum of the 2GI diagrams in Fig.~1 at arbitrary $T$ and $\mu$ can be directly
extracted from from Ref.~\cite{avpres} with the result
\ba
p_\rmi{2GI}
&=&\pi^2 d_A T^4 \Bigg(\fr{1}{45}\bigg\{1+\fr{d_F}{d_A}\left(\fr{7}{4}+30\mubar^2+60\mubar^4\right)\bigg\}\nonumber
\ea
\ba
&-&\fr{g^2}{9(4\pi)^2}\bigg\{C_A + \fr{T_F}{2}(1+12\mubar^2)(5+12\mubar^2)\bigg\}\nn
&-&\fr{g^4}{54(4\pi)^4}\Bigg\{\fr{23C_A^2-C_AT_F(29+360\mubar^2+720\mubar^4) +
4T_F^2(5+72\mubar^2+144\mubar^4)}{\e}\nn
&+&C_A^2\(182\,\ln\fr{\bar{\Lambda}}{4\pi T}+247+272\za-90\zb\)\nn
&+&C_AT_F\bigg(-16\(5+36\mubar^2+72\mubar^4\)\ln\fr{\bar{\Lambda}}{4\pi T}-
\fr{217}{5}-56\za+\fr{72}{5}\zb \nn
&+&24\(9+4\za\)\mubar^2+432\mubar^4+144(1+4\mubar^2)\aleph(1,z)+3456\aleph(3,z)\bigg)\nn
&+&4T_F^2\bigg((1+12\mubar^2)\(4(5+12\mubar^2)\ln\fr{\bar{\Lambda}}{4\pi T}+15+8\za+36\mubar^2\)
\label{2pi}\nn
&+&144(1+4\mubar^2)\aleph(1,z)\bigg)
-9C_FT_F\bigg(\fr{35}{2}-16\za+4\(59+16\za\)\mubar^2\nn
&+&664\mubar^4+96\(i\mubar(1+4\mubar^2)\aleph(0,z)+2(1+8\mubar^2)\aleph(1,z)-12i\mubar\aleph(2,z)\)
\bigg)\Bigg\}\Bigg),
\ea
where $\mubar\equiv
\mu/(2\pi T)$ and where we have
renormalized the gauge coupling using the usual zero-temperature renormalization constant
$Z_g$. The sum, however, still contains uncanceled UV $1/\epsilon$ divergences and depends on the choice of gauge, so that it
has no separate physical significance.

\begin{figure}[t]
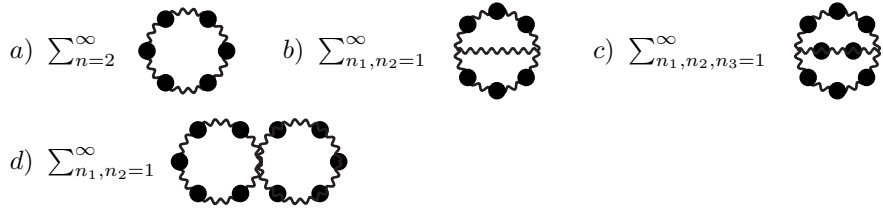

\centering
\ba \nonumber
\begin{array}{lll}
\nn\nn
a)~\sum_{n=2}^{\infty}\!\!\picb{\Agl(15,15)(15,0,360)\GCirc(0,15){2.8}{0.0} \GCirc(7,27){2.8}{0.0} \GCirc(7,3){2.8}{0.0}\GCirc(23,27){2.8}{0.0} \GCirc(23,3){2.8}{0.0}
\GCirc(30,15){2.8}{0.0}}
\;\;\;\;b)~\sum_{n_1,n_2=1}^{\infty}\!\!\picb{\Agl(15,15)(15,0,360)\GCirc(4,25){2.8}{0.0} \GCirc(4,5){2.8}{0.0}\GCirc(15,30){2.8}{0.0} \GCirc(15,0){2.8}{0.0}
\GCirc(26,25){2.8}{0.0} \GCirc(26,5){2.8}{0.0}\Lgl(0,15)(30,15)}
\;\;\;\;c)~\sum_{n_1,n_2,n_3=1}^{\infty}\!\!\picb{\Agl(15,15)(15,0,360)\GCirc(4,25){2.8}{0.0} \GCirc(4,5){2.8}{0.0}\GCirc(15,30){2.8}{0.0} \GCirc(15,0){2.8}{0.0}
\GCirc(26,25){2.8}{0.0} \GCirc(26,5){2.8}{0.0}\GCirc(9,15){2.8}{0.0}\GCirc(21,15){2.8}{0.0}\Lgl(0,15)(30,15)}\nn\nn
d)~\sum_{n_1,n_2=1}^{\infty} \;\, \piccb{\Agl(15,15)(15,0,360)\GCirc(0,15){2.8}{0.0} \GCirc(7,27){2.8}{0.0} \GCirc(7,3){2.8}{0.0}\GCirc(23,27){2.8}{0.0}
\GCirc(23,3){2.8}{0.0}\GCirc(60,15){2.8}{0.0} \GCirc(53,27){2.8}{0.0} \GCirc(53,3){2.8}{0.0}\GCirc(37,27){2.8}{0.0} \GCirc(37,3){2.8}{0.0}\Agl(45,15)(15,0,360)}

\end{array}
\ea
\caption[a]{Classes of IR sensitive vacuum graphs contributing to the QCD pressure at order $g^4$.
The black dots represent the one-loop gluon polarization tensor given in Fig.~3a
and the indices $n_i$ stand for the numbers of loop insertions on the respective lines.}
\end{figure}

\subsection{The ring sum}

To order $g^4$,
the ring sum of Fig.~2a can be separated into three pieces
$p_\rmi{VV}$, $p_\rmi{VM}$ and $p_\rmi{ring}$
according to Fig.~3 by decomposing the one-loop gluon polarization tensor
(see Appendix \ref{app:Pi}) into its vacuum ($T=\mu =0$)
and matter parts. Note that only the matter part has to be resummed, as the vacuum
parts contribute to order $g^4$ only through the two three-loop diagrams
in Figs.~3b and c.%
\footnote{Take any graph $G$ belonging to the ring sum and having four or more loops
and at least one vacuum tensor insertion,
and consider it in the Feynman gauge. Applying Eq.~(\ref{polarvac}) to it and contracting
the Lorentz indices of the vacuum tensor with one of its neighboring gluon
propagators, we see that $G$ is proportional to $g^2$ times a similar graph with the
vacuum insertion removed. But this graph is nothing but one of those diagrams
that appeared in the original sum which implies that $G$ has to be proportional to
at least the fifth power of the coupling. \label{footnoteG}}
The evaluation of $p_\rmi{VV}$ and $p_\rmi{VM}$ is relatively
straightforward, and
fully analytic expressions for them 
are given in Appendix \ref{sec:anlI}.

To evaluate the remaining matter ring sum $p_\rmi{ring}$ we define the standard
longitudinal and transverse parts of the vacuum-subtracted polarization tensor at $d=4-2\e$ by
\ba
\Pi_\rmi{L}(P)\delta^{ab}&=&\fr{P^2}{p^2}\(\Pi_{00}^{ab}(P)-\Pi_{00}^{ab}(P)\mid_{\rmi{vac}}\), \label{pil1}\\
\Pi_\rmi{T}(P)\delta^{ab}&=&\fr{1}{d-2}\(\Pi_{\mu\mu}^{ab}(P)-\Pi_{\mu\mu}^{ab}(P)\mid_{\rmi{vac}} -
\fr{P^2}{p^2}\(\Pi_{00}^{ab}(P)-\Pi_{00}^{ab}(P)\mid_{\rmi{vac}}\)\),\quad\label{pit1}
\ea
where we have used the fact \cite{Heinz:1986kz} that the one-loop
gluon polarization tensor is transverse
with respect to the four-momentum $P$ in the Feynman gauge.
In terms of $\Pi_\rmi{T}$ and $\Pi_\rmi{L}$,
the sum of the ring diagrams is then readily performed with the result
\ba
p_\rmi{ring}&=&-\fr{d_A}{2}\sumint_P\bigg\{\ln\Big[1+\Pi_\rmi{L}(P)/P^2\Big]-\Pi_\rmi{L}(P)/P^2\nn
&+&(d-2)\(\,\ln\Big[1+\Pi_\rmi{T}(P)/P^2\Big]-\,\Pi_\rmi{T}(P)/P^2\)\bigg\}, \label{logsi}
\ea
which is now explicitly IR safe.
\begin{figure}[t]
\centering
\ba \nonumber
\begin{array}{lll}
a)\pic{\Lgl(0,15)(10,15)%
 \Asc(20,15)(10,0,180) \Asc(20,15)(10,180,360) \Lgl(30,15)(40,15)}\!+\pic{\Lgl(0,15)(10,15)%
 \Agh(20,15)(10,0,180) \Agh(20,15)(10,180,360) \Lgl(30,15)(40,15)}\!+\pic{\Lgl(0,15)(10,15)%
 \Agl(20,15)(10,0,180) \Agl(20,15)(10,180,360) \Lgl(30,15)(40,15)}\!+\pic{\Lgl(0,10)(20,10)%
 \Agl(20,20)(10,0,360) \Lgl(20,10)(40,10)}\equiv\, \pic{\Lgl(0,15)(10,15)%
\Lgl(30,15)(40,15)\GCirc(20,15){10}{0.8} \Text(20,15)[c]{V}} \!\!+
\pic{\Lgl(0,15)(10,15)\GCirc(20,15){10}{0.8}\Text(20,15)[c]{M}%
 \Lgl(30,15)(40,15)}\nn\nn
 b)~ p_\rmi{VV}\;\equiv\;\picb{\Agl(15,15)(15,30,150)%
\Agl(15,15)(15,210,330) \GCirc(0,15){7.5}{0.8}\Text(0,15)[c]{V} %
\GCirc(30,15){7.5}{0.8}\Text(30,15)[c]{V}}
\;\;\;\;\;\;
c)~ p_\rmi{VM}\;\equiv\;\picb{\Agl(15,15)(15,30,150)%
\Agl(15,15)(15,210,330) \GCirc(0,15){7.5}{0.8}\Text(0,15)[c]{V} %
\GCirc(30,15){7.5}{0.8}\Text(30.2,15)[c]{M}}
\;\;\;\;\;\;
d)~ p_\rmi{ring}\;\equiv\;\sum_{n=2}^{\infty}\picb{\Agl(15,15)(15,110,160)%
\Agl(15,15)(15,200,250) %
\GCirc(0,15){5}{0.8} \GCirc(15,30){5}{0.8} \GCirc(15,0){5}{0.8} \Agl(15,15)(15,40,70)\Agl(15,15)(15,290,320)
\DAsc(15,15)(15,-40,40)\Text(15,0)[c]{{${\mbox{\scriptsize{M}}}$}}
\Text(15,30)[c]{{${\mbox{\scriptsize{M}}}$}}\Text(0,15)[c]{{${\mbox{\scriptsize{M}}}$}} }
\end{array}
\ea
\caption[a]{a) The one-loop gluon polarization tensor $\Pi_{\mu\nu}(P)$ divided
into its vacuum ($T=\mu=0$) and matter (vacuum-subtracted) parts. \\
b) The IR-safe Vac-Vac diagram contributing to the pressure at $\mathcal{O}(g^4)$. \\
d) The IR-safe Vac-Mat diagram contributing to the pressure at $\mathcal{O}(g^4)$. \\
d) The remaining 'matter' ring sum.}
\end{figure}

As the functions $\Pi_\rmi{L}(P)$ and $\Pi_\rmi{T}(P)$ behave at large $P^2$ like
(see Sec.~B.1.2 of Ref.~\cite{avthesis})
\ba
\Pi_\rmi{L/T}(P)&\xrightarrow[P^2\rightarrow\infty]{}&-2(1+\e)C_Ag^2\sumint_Q \fr{1}{Q^2}
+\mathcal{O}(1/P^2)
\;\;\,\equiv\;\;\,\Pi_\rmi{UV}+\mathcal{O}(1/P^2),
\ea
it is, however, immediately obvious that the sum-integral of Eq.~(\ref{logsi}) is still
logarithmically divergent in the ultraviolet at $T\neq 0$. To regulate the divergence, we add
and subtract a term of the form $(1+d-2)(\Pi_\rmi{UV})^2/(2(P^2+m^2)^2)$ from the integrand, with
$m$ being an arbitrary mass parameter shielding it from IR divergences. By further adding and subtracting
the corresponding massless term from the counterterm, we obtain three separate contributions to $p_\rmi{ring}$:
an UV and IR finite (at least to order $g^4$ --- see below), $m$-dependent ring sum
$p_\rmi{ring}^\rmi{finite}$, an UV finite, but IR divergent and $m$-dependent $p_\rmi{ring}^\rmi{IR}$ and an
UV and IR divergent and massless $p_\rmi{ring}^\rmi{UV}$
\ba
p_\rmi{ring}&=&p_\rmi{ring}^\rmi{finite}+p_\rmi{ring}^\rmi{IR}+p_\rmi{ring}^\rmi{UV},\nonumber
\ea
\ba
p_\rmi{ring}^\rmi{finite}&=&-\fr{d_A}{2}\sumint_P\bigg\{\ln\Big[1+\Pi_\rmi{L}(P)/P^2\Big]-
\Pi_\rmi{L}(P)/P^2+C_A^2g^4T^4/(72(P^2+m^2)^2)\label{prf}\nn
&+&2\(\,\ln\Big[1+\Pi_\rmi{T}(P)/P^2\Big]-\,\Pi_\rmi{T}(P)/P^2+C_A^2g^4T^4/(72(P^2+m^2)^2)\)\bigg\},
\label{eq:pringfinite}\\
p_\rmi{ring}^\rmi{IR}&\equiv&\fr{d_AC_A^2g^4T^4}{48}\sumint_P\bigg\{\fr{1}{(P^2+m^2)^2}-
\fr{1}{P^4}\bigg\}\label{pring37}\\
p_\rmi{ring}^\rmi{UV}&=&\fr{1}{4}(d-1)d_A\(\Pi_\rmi{UV}\)^2\sumint_P{1\over P^4}.
\label{pUV}
\ea
The two first terms can be
evaluated numerically at $\e=0$ while the third one needs to be regulated with finite $\e$. It is noteworthy that
one can set $\e = 0$ even in the formally divergent $p_\rmi{ring}^\rmi{IR}$ due to the fact
that its IR divergence originates solely from the zeroth Matsubara mode of its second
term which vanishes identically in dimensional regularization. The explicit values of $p_\rmi{ring}^\rmi{IR}$ and $p_\rmi{ring}^\rmi{UV}$
are given in Appendix \ref{app:ringsum}, while the numerical evaluation of
$p_\rmi{ring}^\rmi{finite}$ is the subject of Appendix \ref{app:num}.

\subsection{The double and triple sums}

If the sums in Figs.~2b--d were to start from $n=0$, these multiple resummations would clearly correspond to the dressing of the propagators in three two-loop diagrams
with the one-loop gluon polarization tensor. In the present case, we instead define a four-dimensionally transverse\footnote{Thus decomposable into three-dimensionally
transverse and longitudinal parts.} tensor $\Delta G_{\mu\nu}(P)$ by the equations
\ba
\Delta G_\rmi{L}(P)&=&
\fr{1}{P^2+\Pi_\rmi{L}(P)}-\fr{1}{P^2}=
-\fr{\Pi_\rmi{L}(P)}{P^2(P^2+\Pi_\rmi{L}(P))},\\
\Delta G_\rmi{T}(P)&=&
\fr{1}{P^2+\Pi_\rmi{T}(P)}-\fr{1}{P^2}=
-\fr{\Pi_\rmi{T}(P)}{P^2(P^2+\Pi_\rmi{T}(P))},
\ea
corresponding to the difference of a dressed (with the vacuum-subtracted self energy) and a bare gluon propagator in the Feynman gauge. It is a straightforward
exercise in combinatorics to show that the symmetry factors of all graphs in Figs.~2b--d equal $1/4$ independently of $n$ --- a result particularly obvious in
2PI formalism. To order $g^4$, these three classes of diagrams, denoted here by $p_\rmi{b}$, $p_\rmi{c}$ and $p_\rmi{d}$, can then be written in the forms
\ba
p_\rmi{b}&=&\fr{d_AC_A}{4}g^2\sumint_{PQ}\fr{\Delta G_{\mu\mu '}(P)\Delta G_{\rho\rho '}(Q)}{(P+Q)^2}\nn
&\times&\(g^{\mu\nu}(2P+Q)^{\rho}-g^{\nu\rho}(2Q+P)^{\mu}+g^{\rho\mu}(Q-P)^{\nu}\)\nn
&\times&\(g^{\mu '\nu}(2P+Q)^{\rho '}-g^{\nu\rho '}(2Q+P)^{\mu '}+g^{\rho '\mu '}(Q-P)^{\nu}\) + {\mathcal O}(g^6),\label{pb}\\
p_\rmi{c}&=&\fr{d_AC_A}{12}g^2\sumint_{PQ}\Delta G_{\mu\mu '}(P)\Delta G_{\rho\rho '}(Q)\Delta G_{\nu\nu '}(P+Q)\nn
&\times&\(g^{\mu\nu}(2P+Q)^{\rho}-g^{\nu\rho}(2Q+P)^{\mu}+g^{\rho\mu}(Q-P)^{\nu}\)\nn
&\times&\(g^{\mu '\nu '}(2P+Q)^{\rho '}-g^{\nu '\rho '}(2Q+P)^{\mu '}+g^{\rho '\mu '}(Q-P)^{\nu '}\)+ {\mathcal O}(g^6),\\
p_\rmi{d}&=&-\fr{d_A C_A}{2}g^2\sumint_{PQ}\(\Delta G_{\mu\mu}(P)\Delta G_{\nu\nu}(Q)-\Delta G_{\mu\nu}(P)\Delta G_{\mu\nu}(Q)\)+ {\mathcal O}(g^6).\label{pd}
\ea
All contributions involving the vacuum piece of the polarization tensor have been discarded as being of order $g^6$, following a reasoning similar to that in
Footnote~\ref{footnoteG}.

It is worthwhile to first perform a power counting analysis to determine at which order the above sum-integrals start to contribute to the pressure.
In the regime of dimensional reduction, where $T\sim g^x \mu$ with $x<1$, one merely needs to
consider the contributions of the zeroth Matsubara modes, as for the others the temperature acts as an infrared cutoff, leading to their values being proportional
to at least $g^5T\mu^3\sim g^{5+x}\mu^4$. In the region $T\sim g\mu$, the Debye mass is, however, of the same order as the temperature,
implying that all Matsubara modes give contributions to the pressure parametrically similar in magnitude. Scaling the three momenta in the integrals of
Eqs.~(\ref{pb})--(\ref{pd}) by $g\mu$, one quickly sees that the results for the sum-integrals in this regime can up to ${\mathcal O}(g^4)$ be written in the form
$g^4T^2\mu^2f(T/(g\mu))$, where the contributions of the non-static modes to the function $f$ vanish as the parameter $T/(g\mu)$
approaches infinity, while in the opposite limit $T/(g\mu)\rightarrow 0$ the function approaches a constant. As long as we are interested in the value of the pressure
only to order $g^4\mu^4$, these graphs can clearly be altogether
ignored. They will become relevant in the determination of the ${\mathcal O}(g^4\mu^2 T)$ contributions to the specific heats, but this is outside the scope of the
present work.

For now, we can concentrate our attention to the regime of dimensional reduction and therefore to the zero Matsubara mode parts of the above sum-integrals.
Here, we encounter an important simplification which results from the fact that only the longitudinal part of the static gluon polarization tensor has a
non-zero zero momentum limit at one-loop order. As the finite momentum corrections to the functions $\Pi_\rmi{L}(P)$ and $\Pi_\rmi{T}(P)$ clearly correspond
to higher perturbative orders, we can simply replace
\ba
\Delta G_{\mu\nu}(P)&\rightarrow & -\fr{m_\rmi{D}^2}{p^2(p^2+m_\rmi{D}^2)}\delta_{\mu 0}\delta_{\nu 0} \label{glimit}
\ea
in the integrals, leading to a dramatic reduction: both $p_\rmi{c}$ and $p_\rmi{d}$ then vanish identically. This can, however, be easily understood from the point of
view of the three-dimensional effective theory EQCD as a demonstration of the
fact that the $A_0^3$ and $A_0^4$ operators in its Lagrangian
are not accompanied by couplings of order $g$ and $g^2$, respectively,
but only $g^3$ (at nonzero $\mu$) and $g^4$.

In contrast to the above, for $p_\rmi{b}$ one does obtain a non-zero value which has a direct parallel in EQCD in
the form of an ${\mathcal O}(g)$ coupling between one massless $A_i$ and two massive $A_0$ fields and a corresponding two-loop diagram with one $A_i$ and two $A_0$
lines.
Applying the limit of Eq.~(\ref{glimit}) to the sum-integral of Eq.~(\ref{pb}), it is easy to see that we can reduce the expression of $p_\rmi{b}$ (to order $g^4$) to the
simple form
\ba
p_\rmi{b}\;\,=\;\,\fr{d_AC_A}{4}T^2m_\rmi{D}^4g^2\!\int\!\fr{d^3p}{(2\pi)^3}\!\int\!\fr{d^3q}{(2\pi)^3}
\fr{(\mathbf{p}-\mathbf{q})^2}{\mathbf{p}^2(\mathbf{p}^2+m_\rmi{D}^2)\mathbf{q}^2(\mathbf{q}^2+m_\rmi{D}^2)(\mathbf{p}+\mathbf{q})^2}
\ea
which can be solved straightforwardly by introducing three Feynman parameters and using standard formulae for one-loop integrals in three dimensions.
After some work, we get
\ba
p_\rmi{b}&=&\fr{d_AC_A}{4}T^2m_\rmi{D}^4g^2\bigg\{\!\int\!\fr{d^3p}{(2\pi)^3}\fr{1}{\mathbf{p}^2(\mathbf{p}^2+m_\rmi{D}^2)}\int_0^1\!\!\! dx\!\!\int\!\fr{d^3q}{(2\pi)^3}
\fr{1}{\mathbf{q}^2+2x\mathbf{q}\cdot \mathbf{p}+x\mathbf{p}^2+ (1-x)m_\rmi{D}^2}\nn
&-&\(\int_0^1\!\!\! dx\!\!\int\!\fr{d^3p}{(2\pi)^3}\fr{1}{(\mathbf{p}^2+xm_\rmi{D}^2)^2}\)^2\bigg\}\nn
&=&\fr{d_AC_A}{4}\fr{T^2m_\rmi{D}^4g^2}{(4\pi m_\rmi{D})^2}\bigg\{\fr{1}{\pi}\int_0^1\!\!\! dx\fr{1}{\sqrt{x(1-x)}}\!\!\int_0^1\!\!\! dy\fr{1}{\sqrt{y}}\fr{1}{1-y+y/x}
\!\!\int_0^1\!\!\! dz\fr{1}{\sqrt{z}}-1\bigg\}\nn
&=&-\fr{d_AC_A}{4}T^2m_\rmi{D}^2\fr{g^2}{(4\pi)^2}(1-4\,\ln\,2)\label{pbres}
\ea
which we identify as the entire contribution of the classes b-d of Fig.~2 to the QCD pressure up to order $g^4$.

\subsection{The result}\label{subsec:result}

We are now ready to write down our final result for the pressure,
valid on the entire deconfined phase of QCD
and accurate up to and including order $g^4$.
Assembling all the various pieces, this function reads
\ba
p&=&(p_\rmi{2GI}+p_\rmi{VV}+p_\rmi{VM}+p_\rmi{ring}^\rmi{UV}+ p_\rmi{b}) + (p_\rmi{ring}^\rmi{IR} + p_\rmi{ring}^\rmi{finite})
+{\mathcal O}(g^5T\mu^3) +
{\mathcal O}(g^6\mu^4)\label{res1}\\
&\equiv &p_\rmi{anl}+p_\rmi{ring}^\rmi{safe}+{\mathcal O}(g^5T\mu^3) +
{\mathcal O}(g^6\mu^4),
\ea
where $p_\rmi{anl}$ stands for the sum of the first five terms in Eq.~(\ref{res1}) and
\ba
p_\rmi{ring}^\rmi{safe}&\equiv& p_\rmi{ring}^\rmi{finite}+p_\rmi{ring}^\rmi{IR}
\ea
is to be evaluated numerically. One should note that in this notation all $m$-dependence
in contained in the two pieces of $p_\rmi{ring}^\rmi{safe}$, naturally canceling
in their sum. In addition, it is worthwhile to point out that the inclusion of the term $p_b$ in Eq.~(\ref{res1})
is inconsistent in the region of $T\sim g^x\mu$, $x\geq 1$ where we have neglected several
contributions of the same magnitude. As this term, however, is of order $g^{4+2x}\mu^4$, \textit{i.e.~}at least
of order $g^6\mu^4$ in the region in question, the inconsistency is in any case beyond the order to which our result is
indicated to be valid and can therefore be ignored.

Collecting the expressions for all of its parts from above and from
Appendix \ref{app:anl}, the function $p_\rmi{anl}$ reads
\ba
p_\rmi{anl}&=&\pi^2d_AT^4\Bigg(\fr{1}{45}\bigg\{1+\fr{d_F}{d_A}\(\fr{7}{4}+30\mubar^2+60\mubar^4\)\bigg\}\nn
&-&\fr{g^2}{9(4\pi)^2}\bigg\{C_A + \fr{T_F}{2}(1+12\mubar^2)(5+12\mubar^2)\bigg\}\nn
&+&\fr{g^4}{27(4\pi)^4}\bigg\{-C_A^2\bigg(22\,\ln\fr{\bar{\Lambda}}{4\pi T}+63-18\gamma
+110\za-70\zb
\bigg)\nn
&-&C_AT_F\bigg(\(47+792\mubar^2+1584\mubar^4\)\ln\fr{\bar{\Lambda}}{4\pi T}+\fr{2391}{20}+4\za+\fr{116}{5}\zb \nn
&+&6\(257+88\za\)\mubar^2+2220\mubar^4+792(1+4\mubar^2)\aleph(1,z)+3168\aleph(3,z)\bigg)\nn
&+&T_F^2\bigg((1+12\mubar^2)\(4(5+12\mubar^2)\ln\fr{\bar{\Lambda}}{4\pi T}+16\za\)+\fr{99}{5} + \fr{16}{5}\zb\nn
&+&312\mubar^2+624\mubar^4+288(1+4\mubar^2)\aleph(1,z)+1152\aleph(3,z)\bigg)\nn
&+&\fr{9}{4}C_FT_F\bigg(35-32(1-4\mubar^2)\za+472\mubar^2+1328\mubar^4\nn
&+&192\(i\mubar(1+4\mubar^2)\aleph(0,z)+2(1+8\mubar^2)\aleph(1,z)-12i\mubar\aleph(2,z)\)\bigg)\bigg\}\Bigg). \label{panl}
\ea
Not only have all the UV divergences canceled between the
different parts of this result,
once the renormalization of the gauge coupling $g$ has been taken care of,
but this expression actually contains all the (explicit)
renormalization scale dependence
of the pressure up to the present order in perfect agreement with Ref.~\cite{avpres},
leaving $p_\rmi{ring}^\rmi{safe}$ entirely independent of the parameter $\bar{\Lambda}$.
Eq.~(\ref{panl}) is also valid for all values of $T$ and $\mu$; the limit for $\mu\to0$ is given in
Eq.~\nr{mutozero} and the limit $T\rightarrow0$ in Eq.~\nr{ttozero}. All terms non-analytic in $g^2$
are contained in the piece $p_\rmi{ring}^\rmi{safe}$ awaiting numerical evaluation.

In the following we shall denote our final result for the pressure
--- which is accurate to order $g^4$ for all values of $T$ and $\mu$
(while also containing some incomplete contributions of higher order, to be discarded later) --- by
\be
p_\rmi{IV}=p_\rmi{anl}+p_\rmi{ring}^\rmi{safe}.
\ee

\subsection{Numerical infrared issues}\label{sec:numiss}

Before moving on to examining our result by numerically evaluating the function $p_\rmi{ring}^\rmi{finite}$ in Eq.~(\ref{eq:pringfinite}),
there is one more practical issue related to the magnetic
mass problem \cite{Linde:1980ts,Kalashnikov:1980tk} that needs to be dealt with. To wit, in the limit $P\rightarrow 0$, the argument of
$\ln(1+\Pi_\rmi{T}(P)/P^2)$ becomes negative, resulting in an unwanted imaginary
contribution to the integral which actually renders $p_\rmi{ring}^\rmi{finite}$
infrared singular beyond order $g^4$. This problem depends on the choice of gauge, but is present in all
covariant gauges (as well as the Coulomb gauge).

The origin of the problem can be traced back to the fact that when dressed with
the full one-loop self-energy,
the transverse part of the gluon propagator develops a space-like pole.
For $p_0=0$ this pole is determined by the equation \cite{Kalashnikov:1980tk}
\ba
p^2+\Pi_\rmi{T}(p_0=0,p)&=&p^2-g^2N_cT{8+(\xi+1)^2\064}p
\ea
where $\xi$ is the gauge parameter of covariant gauges.\footnote{Replacing the ordinary one-loop gluon self-energy by one that includes
resummation of the Debye mass does not cure the problem, but only produces a different
gauge-dependent spacelike pole \cite{Kalashnikov:1982sc,Kraemmer:2003gd}.}
It is evidently unphysical and appears only at the non-perturbative magnetic mass scale $g^2T$,
which contributes to the pressure starting at order $g^6T^4$. This suggests that
we can in fact eliminate the entire problem by adding by hand
a magnetic mass term
to the transverse self-energy in Eq.~(\ref{eq:pringfinite})
\ba
\Pi_\rmi{T}(P) & \rightarrow & \Pi_\rmi{T}(P) + \mmag^2
\ea
with (for $\xi=1$)
\begin{equation}\label{mmagf}
\mmag=c_{f}\frac{3}{32}g^{2}C_A T
\end{equation}
and $c_f\ge1$,
which only has an effect on the pressure beyond $O(g^4)$.
Indeed, comparing with the effective magnetic mass for nonzero frequencies,
Eq.~(\ref{mmdyn}), we find that the magnetic screening behaviour
is modified only for frequencies $p_0 \lesssim g^4 T$ when $\mu\sim T$
and even $p_0 \lesssim g^4 T(T^2/\mu^2)$ when $T\ll \mu$.
Note, however, that the introduction of this magnetic mass for the transverse self-energy
alters the UV behavior of $p_\rmi{ring}^\rmi{finite}$, implying that both
$p_\rmi{ring}^\rmi{finite}$ and $p_\rmi{ring}^\rmi{IR}$ have to be modified to account
for this reorganization. In
$p_\rmi{ring}^\rmi{finite}$, this change is crucial because it renders the result finite,
but for the already finite $p_\rmi{ring}^\rmi{IR}$ the effects are beyond the order of interest
(see App.~\ref{app:num}).

The numerical evaluation of $p_\rmi{ring}^\rmi{finite}$ is performed along the lines of
Refs.~\cite{moore,ippreb}, with the sum over Matsubara frequencies being converted to an integration
in the usual way (see \textit{e.g.~}Ref.~\cite{kap}). Contributions containing the bosonic
distribution function $n_b$ are best evaluated in Minkowski space, as UV problems are cut off
by $n_b$, while the other contributions are evaluated in Euclidean space in order to
numerically exploit the Euclidean invariance of UV contributions.
By varying the parameter $c_f$ in Eq.~(\ref{mmagf}), we can verify that the effects of this
infrared regulator are indeed beyond the order $g^4$ we are aiming at.
The remaining part,
however, gives rise to yet another type of unphysical pole,
which (at least in the long-wavelength limit) has been well-known since the earliest
perturbative calculations in finite-temperature QCD \cite{Kalashnikov:1979cy}:
in covariant gauges, the one-loop gluon self-energy, evaluated at the location of the
poles corresponding to time-like propagating plasmon modes, gives rise
to a (gauge-dependent) damping constant $\propto g^2 T$ with negative sign
(for all gauge parameters $\xi$, though not in Coulomb or axial gauges
\cite{Kajantie:1982xx,Heinz:1986kz}). A consistent systematic calculation of the
plasmon damping constant to order $g^2T$ requires the use of a HTL-resummed gluon
self-energy which finally leads to a positive and gauge-independent result
\cite{Braaten:1990it,Kraemmer:2003gd}. The corresponding pole is then on the unphysical
sheet where it would cause no problem for the evaluation of $p_\rmi{ring}^\rmi{finite}$.
With the bare one-loop gluon self-energy appearing in our integrand we, however, have poles
on the physical sheet, connected to the light-cone by a branch cut, and we need to avoid
them by deforming the contour of the numerical integration in
Minkowski space as sketched in Fig.~\ref{fig:analyticstructure}.
The details of this procedure and the entire numerical calculation are described further in Appendix \ref{app:num}.

\section{Numerical results}\label{sec:numres}

Having the result of Eq.~(\ref{res1}) for the QCD pressure now finally at hand, we move on to examine
it numerically by evaluating the function
$p_\rmi{ring}^\rmi{safe}$ using methods reviewed in Appendix \ref{app:num} and adding to it the analytic
part of Eq.~(\ref{panl}).
The sum total we call $p_\rmi{IV}$ as a reminder that its
accuracy is of order $g^4$ for all $T$ and $\mu$, while it also includes
incomplete and gauge dependent
higher-order contributions. For the most part of the following analysis, we shall
explicitly eliminate the latter effects by either considering sufficiently small values of $g$
or performing numerical expansions of our results in powers of $g$.

We begin by inspecting
the region where the temperature is parametrically larger than the Debye scale and the
results of dimensional reduction should be applicable,
then continue towards making contact with HDL results
on non-Fermi liquid behavior at $T \lesssim m_\rmi{D}$, and
finally the Freedman-McLerran result for $T\to0$.
In all plots of the present section we use the values
$N_\rmi{c}=3$, $N_\rmi{f}=2$. Because of the latter, we conveniently
have $T_F=1$ and therefore $\tau=\bar\tau$ for the reduced
temperature variables introduced in Eqs.~(\ref{tau}) and (\ref{bartau}), respectively.

\begin{figure}
\begin{center}
\includegraphics[width=10cm]{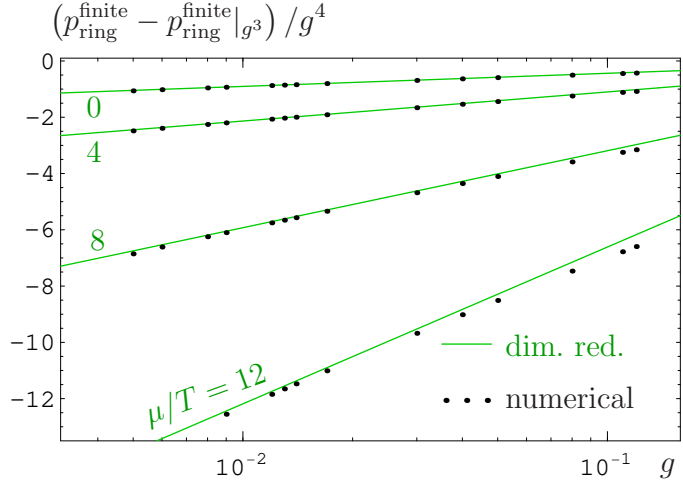}
\caption{Comparison of the $g^4\log g$ and $g^4$ terms of the numerical computation
and the analytic DR result, for various values of $\mu/T$.
The perturbative terms are subtracted up to order $g^3$.
 \label{fig:X}}
\end{center}
\end{figure}

\subsection{$T$ parametrically larger than $m_\rmi{D}$}

The first non-trivial check on our result --- and that of dimensional reduction ---
is to verify that their predictions for the pressure agree to
order $g^4$ for all temperatures and chemical potentials that are of
equal parametric order in $g$.
This is particularly important in order
to clarify that the (entirely correct) statement in the literature about
dimensional reduction being valid as long as $\pi T$ is the
largest dynamical energy scale does not imply a condition
$\pi T > \mu$, but rather $\pi T \gg m_\rmi{D}$ (or even $\pi T \gtrsim m_{\rmi D}$, as
we shall find to be sufficient below).
To this end, we start from the most
widely studied region of $\mu=0$ by comparing our numerical result
to that of the analytic one of dimensional reduction, and then increase the
${\mathcal O}(g^0)$ value of $\mu/T$ up to $\mu \gg \pi T$
while still having $\pi T \gg m_\rmi{D}\sim g\mu$.

The results of this comparison are shown in Fig.~\ref{fig:X}, where we plot the order $g^4\ln\,g$ and $g^4$
contributions of the ring sum of Eq.~(\ref{prf}) to the pressure together with the same quantity
extracted from the result of dimensional reduction (obtained by subtracting the analytic part of our result
from the DR one). The agreement is perfect up to the numerical accuracy of our result, and only at larger values of $g$
can one see that the agreement is getting slightly worse with increasing $\mu/T$. This was, however, to be expected,
since there $\mu/T$ is coming closer to the value $g^{-1}$, making $m_\rmi{D}/T$ of order one which is parametrically
the limit of applicability of dimensional reduction. Our conclusion is that the result of dimensional reduction
is valid at in principle arbitrarily large ${\mathcal O}(g^0)$ values of $\mu/T$, though the expansion in $g$ only
makes sense at smaller and smaller values of $g$ as this parameter is increased. This statement will be made more
concrete in the following sections.

\begin{figure}
\begin{center}
\includegraphics[width=10cm]{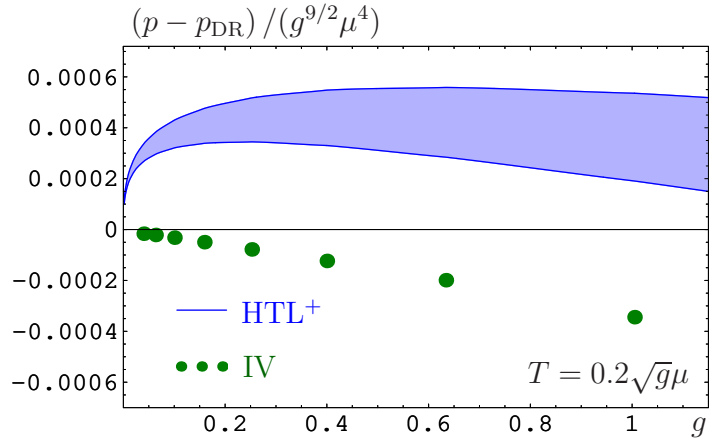}
\caption{Comparison of the HTL+ pressure and our numerical result $p_\rmi{IV}$ in the region of $T=\tau \sqrt{g} \mu$, $\tau=0.2$, with the known
perturbative terms from dimensional reduction
 subtracted and the entire quantities divided by $g^{9/2}$.
This plot shows that both the HTL+ result and our numerical one are accurate
at least up to order $g^{9/2}$. The renormalization scale has
been varied between $\mu$ and $4\mu$. While $p_\rmi{IV}-p_\rmi{DR}$
is scale independent, $p_\rmi{HTL+}-p_\rmi{DR}$ has a scale dependence
at order $g^4\mu^2 T^2\sim g^5\mu^4$.
 \label{fig:ninehalf}}
\end{center}
\end{figure}

\begin{figure}
\begin{center}
\includegraphics[width=10cm]{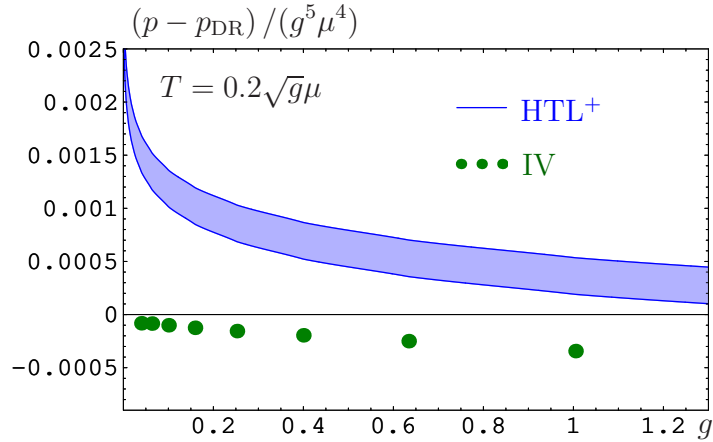}
\caption{Same as Fig.~\ref{fig:ninehalf}, but normalized to $g^5$.
While the HTL+ result is no longer accurate to this order and diverges logarithmically,
our numerical result still correctly reproduces the
dimensional reduction result for the pressure at order $g^5\mu^4$.
 \label{fig:ninehalfb}}
\end{center}
\end{figure}

The logical next step is to test the validity of dimensional reduction at temperatures larger
than but now parametrically closer to the Debye scale.
For concreteness, we specialize to the case of $T\sim \sqrt{g}\mu$, for which the prediction
of dimensional reduction is given in Eq.~(\ref{pDRsqrtg}).
In this region, the error in our result is of order $g^{11/2}\mu^4$ and that of the minimal
HTL resummation $g^5 \mu^4$,
so that the first one should be able to reproduce the first seven and the latter the
first six terms of the series (\ref{pDRsqrtg}). And indeed, a numerical
evaluation of both Eqs.~(\ref{res1}) and (\ref{PHTL}) and the subtraction of
the first terms of Eq.~(\ref{pDRsqrtg}) shows the expected results: as
displayed in Fig.~\ref{fig:ninehalf}, we find perfect agreement in comparing the
dimensional reduction result with the HTL one (\ref{PHTLtot}) and
with that of our new approach up to order $g^{9/2}$.
In Fig.~\ref{fig:ninehalfb}, we see that our numerical evaluation
of $p_\rmi{IV}$ is accurate enough to even
verify the $g^5\mu^4$ term in the dimensional reduction result, while the HTL result starts deviating from the
DR one at this order.

\subsection{$T$ comparable to $m_\rmi{D}$}

\begin{figure}
\begin{center}
\includegraphics[width=10cm]{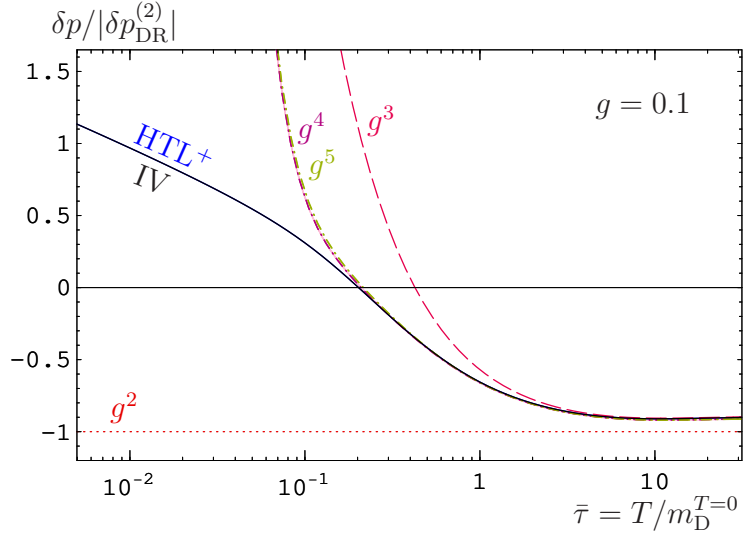}
\caption{
Thermal contribution to the interaction pressure $\delta p$
as a function of $T/m_\rmi{D}^{T=0}$ for fixed chemical potential $\mu$
and coupling $g=0.1$. 
For this value of the coupling, the results of the numerical evaluation
of $p_\rmi{anl}+p_\rmi{ring}^\rmi{safe}$ and $\rm{HTL}^{+}$
coincide within plot resolution.
The result is compared to the dimensional reduction pressure
at orders $g^2$, $g^3$, $g^4$, and $g^5$ (where the latter
is included only for completeness, as neither
$p_\rmi{IV}$ nor $p_{\rmi{HTL}^{+}}$
contain contributions of order $g^5$). The effect of varying the
renormalization scale $\lambdamsbar=\mu \,...\, 4\mu$ is not visible
for this value of the coupling. \label{fig:p01}}
\end{center}
\end{figure}

\begin{figure}
\begin{center}
\includegraphics[width=10cm]{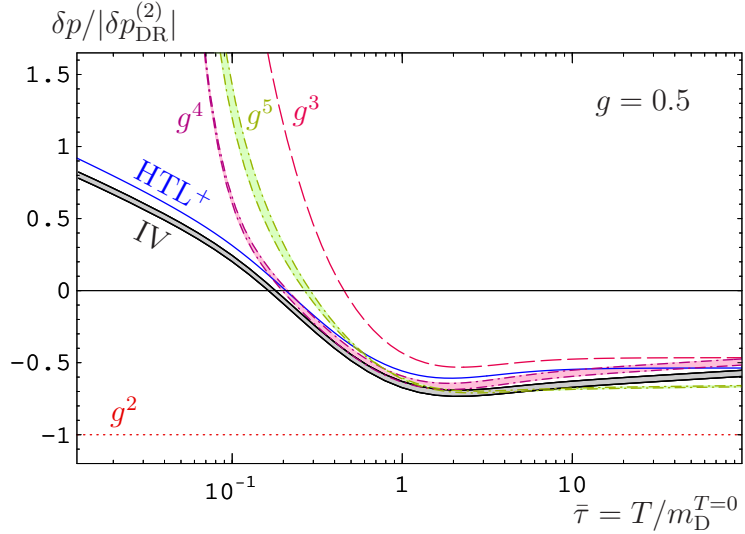}
\caption{
Same as Fig.~\ref{fig:p01}, but for $g=0.5$.
The results of the numerical evaluation
of $p_\rmi{anl}+p_\rmi{ring}^\rmi{safe}$ and $\rmi{HTL}^{+}$
can now be distinguished due to their different content of higher-order terms.
When two lines of the same type run close to each other, they differ by
changing the renormalization scale $\lambdamsbar=\mu \,...\, 4\mu$.
\label{fig:p05}}
\end{center}
\end{figure}

\begin{figure}
\begin{center}
\includegraphics[width=10cm]{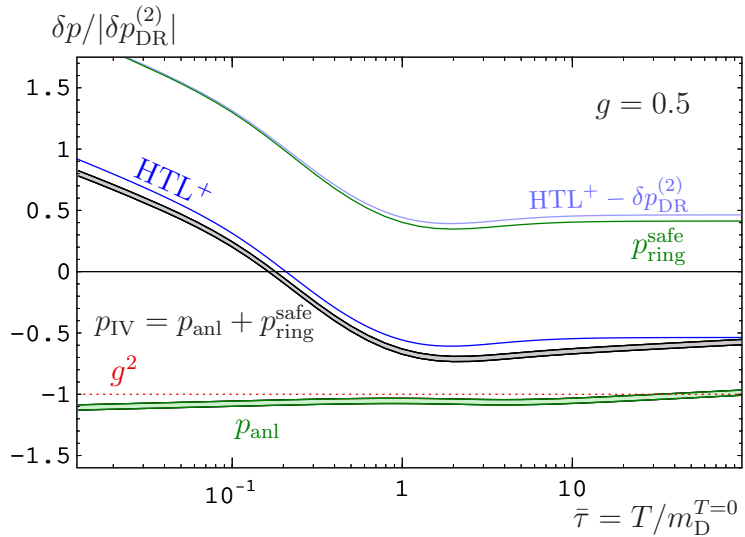}
\caption{
Same as Fig.~\ref{fig:p05}, but with $p_\rmi{IV}$ separated into $p_\rmi{anl}$ and $p_\rmi{ring}^\rmi{safe}$.
As the $g^4$ contribution in $\delta p_\rmi{anl}$
only amounts to a small correction
(of effective order $g^6$),
the shape of the full pressure curve as a function of $T$
(beyond the rather trivial $g^2$ contribution)
is mainly determined by $p_\rmi{ring}^\rmi{safe}$.
The renormalization scale dependence
$\lambdamsbar=\mu\, ...\, 4\mu$ is entirely due to $p_\rmi{anl}$.
\label{fig:p05parts}}
\end{center}
\end{figure}

In Figs.~\ref{fig:p01}--\ref{fig:p05parts}, we plot the the temperature-dependent contributions to the interaction pressure $\delta p$
(see Eq.~(\ref{deltaDeltap})) for $T\sim m_\rmi{D}$ as extracted from our numerical
calculation of $p_\rmi{IV}$ but with no expansions in powers of $g$. We compare this with $p_\rmi{HTL+}$ as well
as with the dimensional reduction result expanded to orders $g^2$, $g^3$, $g^4$ and $g^5$ which refer
to the counting in powers of $g$ when $T\sim\mu$.
For $T\sim g\mu$, however,
the terms $g^2 \mu^2 T^2$, $g^3 \mu^3 T$, and $g^4\mu^4\ln\,T$ all
become of the same order of magnitude and together constitute the leading temperature-dependent
contribution to the interaction pressure $p-p_\rmi{SB}$ which is contained in the result marked by
the dashed line ``$g^4$''. For completeness, we also
include the complete dimensional reduction result to
(explicit) order $g^5$, but it should be remembered that
the term $g^5 T\mu^3$ is already of the same magnitude as the unknown $g^6\mu^4$ piece
when $T\sim g\mu$, and is therefore both incomplete and beyond our scope which also explains why the $g^4$ curve seems to produce better agreement
with our results than the $g^5$ one.

The different results are normalized to the leading term of the $T$-dependent part of the interaction pressure
in the dimensional reduction result
(\ref{dimredpr5}),
\be\label{dpDR2}
\delta p^{(2)}_\rmi{DR}  =  -g^2
d_A \biggl\{ {T_F\016\pi^2} \mu^2 T^2 + {5T_F+2C_A\0244} T^4 \biggr\}.
\ee
To understand the structure of these figures, note that the $g^3$ curve goes like $-1+(4/3\pi) m_\rmi{D}/T$
for small $T$ and like $-1+1.07g$ for large $T$. At $T\ll m_\rmi{D}$ it, of course, deviates from the
exact result which is instead dominated by the
leading $\fr{2}{9} \ln \, T^{-1}$ behavior of the low-temperature series of Eq.~(\ref{PHDLx}) when
normalized by the absolute value of Eq.~\nr{dpDR2}.

For small values of $g\sim 0.1$, Fig.~\ref{fig:p01} shows that the numerical evaluation
of Eq.~(\ref{res1}) perfectly agrees with the result of the
HTL resummation (the two curves lie virtually on top of each other).
At this value of $g$, also the complete dimensional reduction result to
(explicit) order $g^5$ is virtually indistinguishable from the
order $g^4$ result. The dimensional reduction result
reproduces the numerical results remarkably well down to temperatures of about
$0.2\,m_\rmi{D}^{T=0}$, but at even lower $T$ severely overestimates the
logarithmic growth of
$\delta p/T^2$ as $T\to 0$.
This is to be expected, since, in the limit $T\to0$,
the plasmon term of order
$g^3\mu^3 T$ in the pressure is clearly unphysical, as it would
lead to a nonvanishing entropy at $T=0$; the $g^4 \mu^4 \ln\,T$ term
of the dimensional reduction result,
while evidently crucial for good agreement down to $T\approx 0.2 m_\rmi{D}$,
would even lead to a diverging entropy as $T\to0$.
The point at which the dimensional reduction result ceases to be a
good approximation for both $p_\rmi{HTL+}$ and $p_\rmi{IV}$ seems
to agree rather well with the value of $T/m_\rmi{D}$ where $\delta p$
switches sign.

In Fig.~\ref{fig:p05} we consider a larger coupling $g=0.5$, for which we begin to see effects from
varying the renormalization scale $\lambdamsbar$ in our result by a factor of 2 around
the central value $2\mu$,
except in the HTL+ result, where $\lambdamsbar$ appears only
in the $T=0$ (Freedman-McLerran) part of the result.\footnote{
In Fig.~\ref{fig:p05}, the value $g=0.5$ is kept fixed for all $\lambdamsbar$,
which means that the $x$-axis does not correspond to a
renormalization-group invariant variable. The (explicit)
dependence of the results on $\lambdamsbar$ is here shown
only to assess the theoretical error in the numerical comparison between the
different approaches. Taking into account the implicit
$\lambdamsbar$ dependence of $g$, the scale dependence of
all the results we are comparing is of the order of their error,
which is $O(g^6\mu^4)$ at $T\sim g\mu$.}
For small $T/m_\rmi{D}^{T=0}$, we find good agreement between the HTL+ result
and $p_\rmi{IV}$, 
with the dimensional reduction result to order $g^4$ lying in between the two
in the range $T/m_\rmi{D}^{T=0}\approx 0.1\ldots 10$, but deviating again
abruptly for $T/m_\rmi{D}^{T=0} < 0.2$,
which is where $\delta p$ changes sign.
At this value of the coupling, the complete order $g^5$ result
of dimensional reduction is still reasonably close to the order $g^4$
result. While it is certainly unreliable when $\delta p>0$, the
order $g^5$ result suggests that taking into account the next
higher orders in $g$ may move the onset of non-Fermi-liquid behavior
to slightly larger $T/m_\rmi{D}$.

Fig.~\ref{fig:p05parts} shows how the final result $\delta p_\rmi{IV}$
is composed of the infrared-safe piece $p_\rmi{anl}$ and the ring sum
$p_\rmi{ring}^\rmi{safe}$. At parametrically small $T\sim g\mu$,
the $T$-dependent terms in the interaction part of $p_\rmi{anl}$
which are of effective order $g^4 \mu^4$ come just from the terms $g^2 T^2 \mu^2$,
so that the shape of $\delta p$ in the above figures is mainly
determined by $\delta p_\rmi{ring}^\rmi{safe}$ which is seen to
coincide with $\delta p_\rmi{HTL+}-\delta p^{(2)}_\rmi{DR}$ up
to terms beyond $g^4$ accuracy.

\begin{figure}
\begin{center}
\includegraphics[width=10cm]{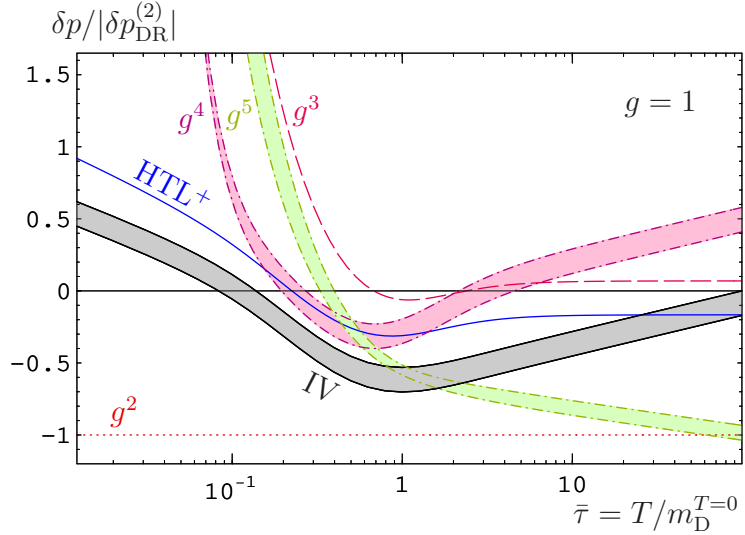}
\caption{
Same as Fig.~\ref{fig:p01}, but for $g=1$.
At this value of the coupling, the numerical result for $p_\rmi{IV}$
begins to be visibly affected by the choice of the
magnetic mass (\ref{mmagf}) which here is taken with $c_f=1$.
\label{fig:p1}}
\end{center}
\end{figure}

Finally, in Fig.~\ref{fig:p1} we consider $g=1$ which is roughly the value
of the QCD coupling at 100 GeV. Here the result for $p_\rmi{IV}$
still follows $p_\rmi{HTL+}$ for $T<m_\rmi{D}$ and the $g^4$ result of
dimensional reduction for $T\gg m_\rmi{D}$, but there is an overall shift
due to the
(incomplete and gauge dependent) order $g^6\mu^4$ terms in $p_\rmi{IV}$.
At this value of $g$ one also notices that the dimensional reduction
result to order $g^4$ deviates rather strongly from the result to order $g^5$,
thus showing poor apparent convergence, in particular at high $\tau$.

\begin{figure}
\begin{center}
\includegraphics[width=10cm]{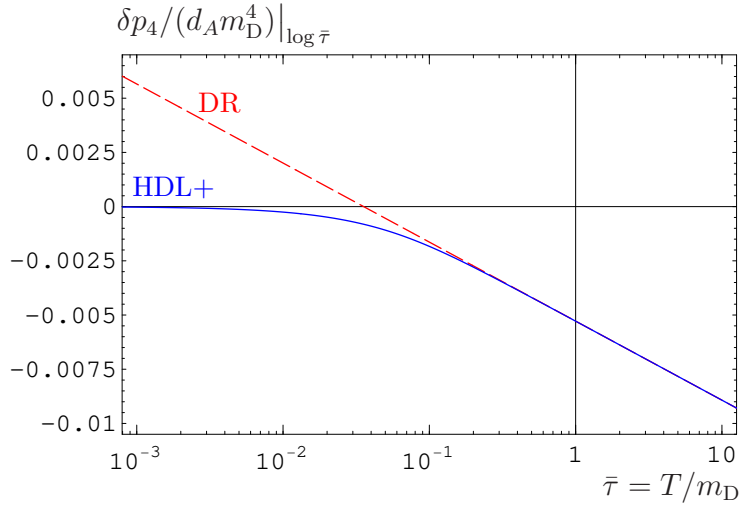}
\caption{Plot of the last term in
Eq.~(\ref{dp4DR}) (dashed line) in comparison with $\delta p_\rmi{HDL$^+$}$
with the first two terms of Eq.~(\ref{dp4DR})
subtracted, in units of $m_\rmi{D}^4=\g^4\mu^4/\pi^4$
and as a function of $\log_{10}(\t)$.}
\label{fig:dpt4}
\end{center}
\end{figure}

For the remainder of the discussion of our numerical results at $T\sim m_\rmi{D}$, we concentrate on the contributions of order $g^4\mu^4$ to the pressure
and explicitly discard all terms beyond this accuracy, as this helps us to better analyze the breakdown of dimensional reduction observed in the
three previous plots. To this end, we consider the
difference of the $g^4\mu^4$ term in the DR result at $T\sim g\mu$ and the corresponding piece in the Freedman-McLerran zero-temperature expression.
A straightforward evaluation of this quantity gives
\ba\label{dp4DR}
{1\0d_A\mu^4}\delta p^\rmi{DR}(\mu,\t={T\over m_\rmi{D}})&=& -{\g^4\t^2\016\pi^4}
+{\g^4\t\012\pi^5}\\
&-&{\g^4\0768\pi^6}\left[
33-3\d-12\gamma-2\pi^2+2\,\ln\,2(8\,\ln\,2+7)+12\,\ln(\pi\t)\right]
+O(g^6)\nonumber
\ea
which, like the HDL$^+$ result, only depends on $g$ and $T$ to order $g^4$
through the combinations $\g\equiv T_F^{1/2}g$ and $\t=\pi T/(\g \mu)=T/m_\rmi{D}$
and where $\delta$ is the numerical constant appearing in Eq.~(\ref{pFMcL}).
As we have seen, in this region the $g^4$ content of our $p_\rmi{IV}$
agrees perfectly with that of $p_\rmi{HDL$^+$}$ defined by Eqs.~(\ref{PHDL})
and (\ref{deltaDeltap}). Therefore, to simplify our numerical efforts, we compare in Fig.~\ref{fig:dpt4}
the above function with the HDL$^+$ result for the same quantity $\delta p$, after by hand subtracting
the terms proportional to $\t^2$ and $\t$ of Eq.~(\ref{dp4DR}) from both results (which correspond to the terms of order
$g^2\mu^2 T^2$ and $g^3\mu^3 T$ in the expansion of the pressure).\footnote{Note that these terms are not present in the HDL$^+$ result at small values of $\t$
(but that this does not matter, as their effect in Fig.~11 in any case vanishes as $\t \to0$).} This helps us to expose the term of order
$g^4\mu^4\ln\,T$, whose divergence at $T\rightarrow 0$ signals the failure of dimensional
reduction to correctly describe the zero-temperature limit of the pressure.

From Fig.~\ref{fig:dpt4}, we observe that for $\t\gtrsim 0.2 m_\rmi{D}$
there is a $\ln\, \t$ term in the HDL$^+$ result which agrees perfectly
with that of Eq.~(\ref{dp4DR}).
This fact turns out to have a natural explanation which gives us important insight
into the breakdown of dimensional reduction.
The key observation is that the ultimate reason for this breakdown lies in the
incorrect treatment of low-temperature IR divergences in DR: in deriving its prediction for the QCD pressure,
one assumes that the temperature acts as the sole IR cut-off for all non-zero bosonic Matsubara frequencies in the logarithmically IR
divergent three-loop ring diagrams, so that for them no resummations are necessary.\footnote{Which amounts to expanding the already resummed propagators
for the non-static gluon modes in these diagrams in powers of $g\mu/T$.} While this indeed is justified for
$T\gg m_{\rmi D}$, there is an obvious problem in the region $T\sim m_{\rmi D}$, ultimately leading to the diverging of the DR result
in the zero-temperature limit where discrete Matsubara modes no longer exist. In
a physically consistent calculation, where one performs a resummation also for the non-static modes, the logarithm of temperature in
the dimensional reduction result above gets replaced by one of a true IR regulator $T\times R(g\mu/T)$, where the function $R$ is linear at large values of its
argument and approaches a non-zero constant as $g\mu/T\to0$. Unlike $T$, this regulator therefore does not vanish at $T=0$ which explains the finite $\t\rightarrow 0$
limit of the HDL$^+$ curve in Fig.~\ref{fig:dpt4}. However, at large values of $\t$ the logarithm of $TR$
gives rise to the $\ln\,\t$ behavior with the same coefficient as in
the dimensional reduction result which is what we observe in Fig.~\ref{fig:dpt4}.

\subsection{$T$ smaller than $m_\rmi{D}$}

\begin{figure}
\begin{center}
\includegraphics[width=10cm]{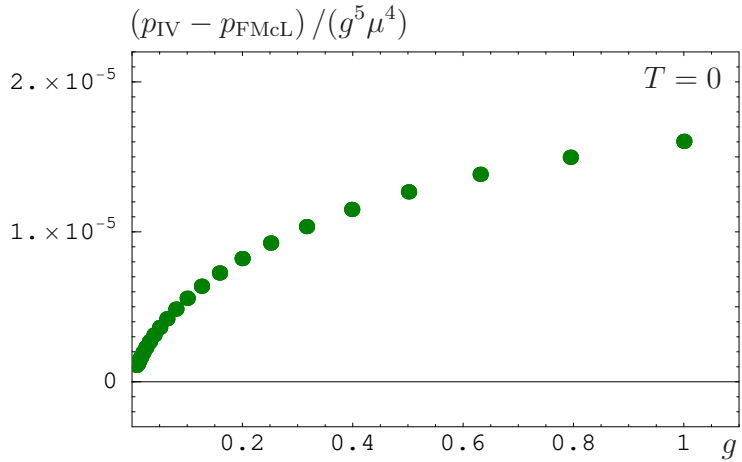}
\caption{Comparison of our numerical result $p_\rmi{IV}$ and the Freedman-McLerran result.
The plot shows agreement to order $g^5$, {\it i.e.,}
agreement in the nonvanishing coefficients up to and including order $g^4$,
and the absence of order $g^5$ contributions.
 \label{fig:FMcLfive}}
\end{center}
\end{figure}

\begin{figure}
\begin{center}
\includegraphics[width=10cm]{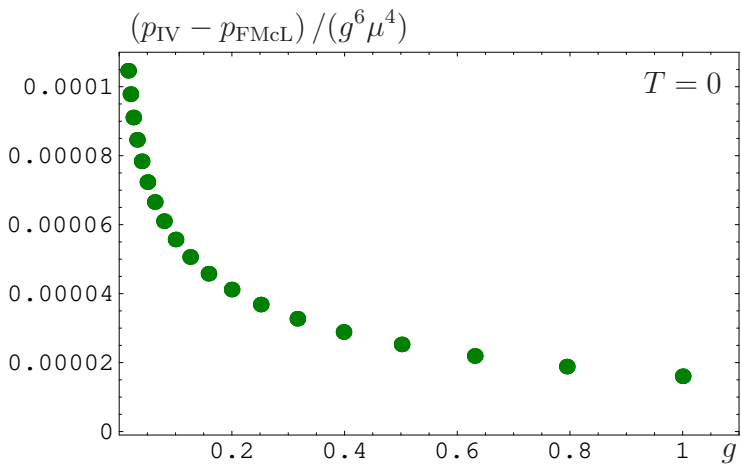}
\caption{Same as Fig.~\ref{fig:FMcLsix}, but divided by $g^6$ instead of
$g^5$, revealing that $p_\rmi{IV}$ contains a term of order $g^6 \ln\,g\, \mu^4$,
which is however incomplete as it is beyond the accuracy of our setup.
 \label{fig:FMcLsix}}
\end{center}
\end{figure}

In the limit $\bar\tau=T/m_\rmi{D}\to 0$, our evaluation of $p_\rmi{IV}$
approaches the Freedman-McLerran result (\ref{pFMcL})
as shown in Figs.~\ref{fig:FMcLfive} and \ref{fig:FMcLsix},
where the difference of $p_\rmi{IV}(T=0)$ and $p_\rmi{FMcL}$ is plotted
as a function of $g$. The agreement to order $g^4$ and the absence
of a $g^5$ term is shown in Fig.~\ref{fig:FMcLfive}, where we normalize
the result by $g^5\mu^4$. Fig.~\ref{fig:FMcLsix} on the other hand shows that
$p_\rmi{IV}$ differs from the Freedman-McLerran result at order $g^6$, where it contains a term
of order $g^6 \ln\,g$. This, however, is incomplete, as the true (unknown) $g^6\ln\,g$ term in
the $T=0$ pressure gets contributions also from the two-loop gluon self energy.

\begin{figure}
\begin{center}
\includegraphics[width=10cm]{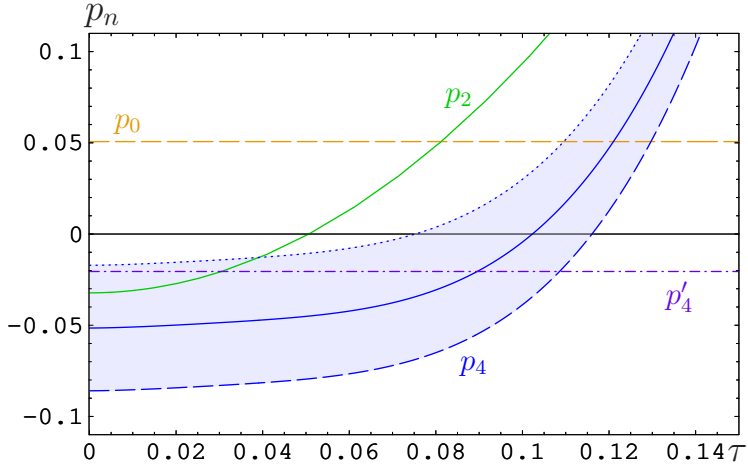}
\caption{
Dimensionless coefficients of the expansion of the pressure
in powers of $g$ at
$\pi T = \tau g \mu$ as a function of $\tau$ for $N_c = 3$, $N_f = 2$.
The coefficients $p_0$ and $p_4'$ are constants, $p_2$ is quadratic in $\tau$, while $p_4$ shows a $\tau^2 \log \tau$ behavior for small $\tau$.
The renormalization scale is varied through $\lambdamsbar=\mu\ldots 4 \mu$.
 \label{fig:pressurecoeffsqcdsmall}}
\end{center}
\end{figure}

In Fig.~\ref{fig:pressurecoeffsqcdsmall} we plot the coefficients
of the low-temperature pressure in the perturbative expansion
\ba
\frac{p}{\mu^4} = p_0 + \frac{g^2}{4\pi}p_2 + \frac{g^4}{(4\pi)^2} \left(p_4 + p'_4 \log g \right)+{\mathcal O}(g^6\ln\,g)
\ea
as a function of $\tau=\pi T/(g\mu)$ for $N_c = 3$ and $N_f = 2$.
Here, we have again numerically evaluated the HDL$^+$ result which
as we have shown agrees with our new approach to order $g^4$ for
$T\sim g\mu$. For $\tau=0$ the coefficients reproduce the
Freedman-McLerran result, whereas for larger $\tau$ the coefficients
are dominated by Stefan-Boltzmann contributions $T^2\mu^2 \sim g^2 \mu^4$
and $T^4\sim g^4 \mu^4$.

\begin{figure}
\begin{center}
\includegraphics[width=10cm]{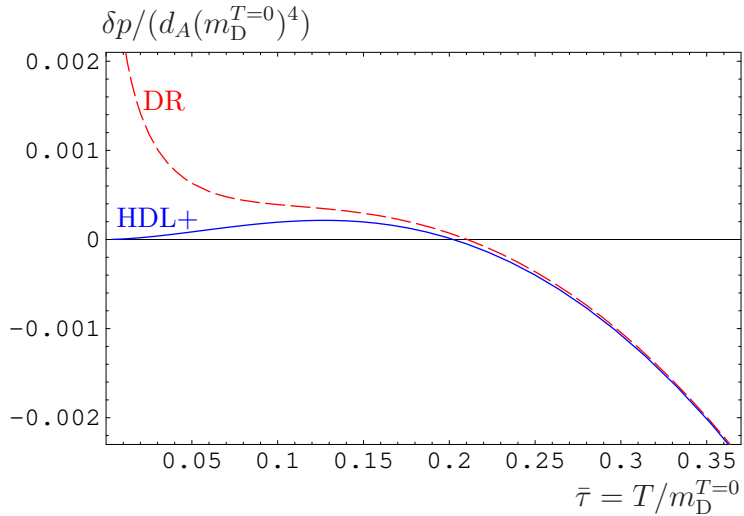}
\caption{Plot of the $T$-dependent part of the interaction pressure
$\delta p$ (see Eq.~(\ref{deltaDeltap})) to order $g^4$
in the regime $\pi T\lesssim m_\rmi{D}$ in units of $(m_\rmi{D}^{T=0})^4=\g^4\mu^4/\pi^4$
and as a function of $\t=T/m_\rmi{D}$. The dashed line denotes the
dimensional reduction result to order $g^4$, and the full line
the HDL$^+$ result which in this regime coincides with the order-$g^4$
content of $p_\rmi{IV}$.}
\label{fig:deltap}
\end{center}
\end{figure}

In Fig.~\ref{fig:deltap}, we compare the
$T$-dependent contribution to the interaction pressure, $\delta p$,
with the extrapolation of the dimensional
reduction result (where $p_\rmi{FMcL}$ has been
subtracted since $p_\rmi{DR}$ does not exist at $T=0$)
to $T\ll m_\rmi{D}$ in a linear plot.
Here, $\delta p$ is normalized by $m_\rmi{D}^4\sim g^4\mu^4$ rather
than divided by a term quadratic in temperature to better show
its absolute magnitude\footnote{The normalization to
$\delta p_\rmi{DR}^{(2)}$ is however more appropriate if one
is interested in the magnitude of the effects to entropy and
specific heat.}.
As could already be seen in the previous plots, the dimensional
reduction result works surprisingly well down to temperatures
where $\delta p$ changes sign which happens at $T\approx 0.2 m_\rmi{D}$.
At even lower temperatures, the dimensional reduction result
overestimates the leading positive contribution to $\delta p$ which
in reality goes to zero like (or grows $\sim\log T$ with the normalization of
Figs.~\ref{fig:p01}--\ref{fig:p05parts})
\ba
\delta p(\mu,T\ll g\mu) &=&
{d_A T_F\072\pi^2}g^2\mu^2T^2 \ln {cT_F^{1/2}g\mu\0T} +
O\left((g\mu)^{4/3}T^{8/3}\right) \label{smallt}
\ea
with the constant $c\approx 0.284794$. This expression
determines the total pressure of the normal QCD phase as
\be
p_\rmi{normal}(\mu,T\ll g\mu)=p_\rmi{normal}(\mu,0)+\delta p(\mu,T\ll g\mu),
\ee
where $p_\rmi{normal}(\mu,0)$ is given in Eq.~(\ref{pFMcL}).

For exponentially small $T$ with $\ln (g\mu/T) \gtrsim g^{-2}$ the
contribution of Eq.~ (\ref{smallt})
can even be larger than the leading $T$-dependent\footnote{Note that in the entropy or in
the specific heat these terms constitute the leading ones.}
term $\propto \mu^2 T^2$
in the Stefan-Boltzmann pressure \cite{Schafer:2004zf}.
However, before one reaches temperatures so small
that $\ln (g\mu/T)\gtrsim g^{-2}$ in a nonabelian plasma,
one encounters the nonperturbative
pair instability of color superconductivity
and the formation of a gap \cite{Son:1998uk}
$\phi \sim T_c \sim \mu
g^{-5} \exp[-3\pi^2/(g \sqrt2)]$ at the parametrically
larger temperatures with $\ln (g\mu/T_c)\gtrsim g^{-1}$.
In our (resummed) perturbative approach, we will not directly
encounter this nonperturbative instability, but since
at the superconducting transition temperature the pressures
of the normal and superconducting
phases in any case have to be equal, it is of some interest
to evaluate the correction of Eq.~(\ref{smallt})
at the transition temperature $T_c^\rmi{SC}$.
Inserting $\ln(T/\mu) \approx -3\pi^2/(g \sqrt2)$ into
Eq.~(\ref{smallt}), one finds
\be
p_\rmi{normal}(\mu,T_c^\rmi{SC})=p_\rmi{normal}(\mu,0)+
{d_AT_F\024\sqrt2}\,g\,\mu^2(T_c^\rmi{SC})^2.
\ee
Comparing this to the
$T=0$ pressure of, {\it e.g.,} the 2SC phase from \cite{Schmitt:2004et}
and expressing it in terms of $T_c$
by replacing $\phi \to \pi e^{-\gamma}T_c^\rmi{SC}$, we have
\be
p_\rmi{SC}(\mu,0)
\approx p_\rmi{normal}(\mu,0)
+e^{-2\gamma}T_F\mu^2
(T_c^\rmi{SC})^2.
\ee
Thus we observe that at the temperature where
color superconductivity sets in, the contribution of resummed perturbation theory is
$g d_A  e^{2\gamma}/ (24 \sqrt{2}) \approx 0.75 g$ times
the $T=0$ contribution of the gap.

\section{Summary of perturbation theory on the $\mu$-$T$ plane}\label{sec:summary}

As we have seen in the previous Section, the weak coupling expansion of the QCD pressure
goes through changes in its form when $T/\mu$ becomes comparable to some positive power of the coupling constant $g$
and this power is then increased. In particular, at $T\lesssim g\mu$ dimensional reduction ceases to be applicable and a resummation of the nonstatic parts of
the gluon self energy becomes necessary, although numerically the dimensional reduction result works surprisingly well
down to the rather small value $T/m_\rmi{D}\approx 0.2$. At this
temperature, which to leading order in the coupling reads
\be\label{TNFL}
\frac{T_{\rm NFL}}{\mu} \approx 0.064 \sqrt{\frac{N_f}{2}}{g},
\ee
the $T$-dependent contributions to the interaction
pressure change sign, marking the onset of so-called non-Fermi-liquid
behavior.
At parametrically even smaller temperatures, one eventually
encounters the critical temperature of color superconductivity,
which has been calculated to leading order in a weak coupling analysis
as \cite{Brown}
\be\label{TCSC}
\frac{T_c^{\rm SC}}{\mu}\simeq
2\frac{e^\gamma}{\pi}
e^{-(\pi^2+4)/8}(4\pi)^4 \left(\frac{2}{N_f}\right)^{5/2} g^{-5} e^{-3\pi^2/\sqrt2 g}
\ee
for a spin-zero condensate (which gives the largest value of $T_c$).

\begin{figure}
\begin{center}
\includegraphics[width=9cm]{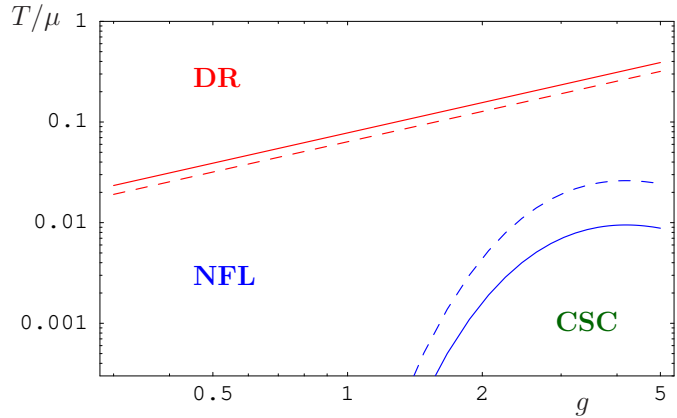}
\caption{The dividing line between the regime of dimensional reduction (DR)
and that of non-Fermi-liquid behavior (NFL) as given by Eq.~(\ref{TNFL})
for $N_f=3$ (full lines) and $N_f=2$ (dashed
lines), in comparison with the
weak-coupling result (\ref{TCSC}) for the
critical temperature of color superconductivity (CSC)
when extrapolated to large coupling. 
\label{fig:Tmu}}
\end{center}
\end{figure}

In Fig.~\ref{fig:Tmu}, we compare $T_{\rm NFL}$ and $T_c^{\rm SC}$
when extrapolated to large coupling (where of course correspondingly
large modifications can be expected). At least for smaller coupling,
i.e., sufficiently high densities, there is a clear separation of 
regimes in the $\mu$-$T$ plane with qualitatively different weak coupling
descriptions.

\begin{figure}
\begin{center}
\includegraphics[width=11cm]{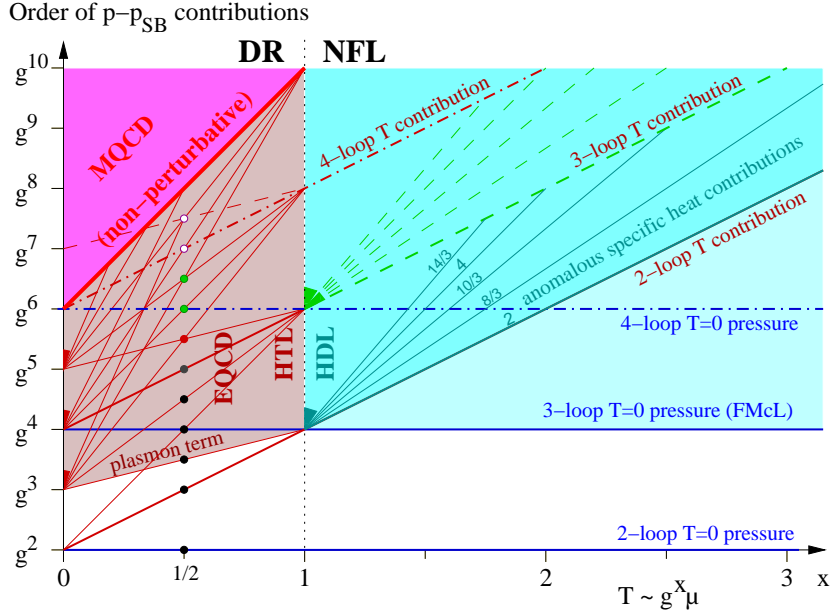}
\caption{The structure of the weak-coupling expansion of the
interaction pressure $p-p_\rmi{SB}$ at parametrically
small $T/\mu$ as a function of the power $x$ in $T\sim g^x \mu$.
At $T\sim\mu$, \textit{i.e.}~$x=0$, the expansion
involves orders 2, 3, 4,\ldots in $g$ (logarithms in $g$ are
not made explicit); at $T\sim g^{1/2}\mu$ where
dimensional reduction overlaps with HTL/HDL resummation,
the series in $g$ involves powers 2, 3, $7\over2$, 4, $9\over2$,\ldots;
at $T\sim g\mu$, where dimensional reduction ceases to
be applicable, the expansion is again in even powers of $g$ (and logs)
with coefficients that at even smaller temperatures
can be expanded in a series involving fractional powers
of $T$ (beginning with
2, $8\over3$, $10\over3$, 4, $14\over3$,\ldots) and corresponding
powers $2+2x, 2+{8\over3}x,\ldots$ of $g$. While subleading in
the pressure, the latter contributions give the leading-order
``anomalous'' (non-Fermi-liquid)
contributions to the interaction part of the entropy and
specific heat at low temperature. Existing results
for the various contributions are represented by full
lines, as-yet undetermined contributions
by dashed and dash-dotted lines. The
non-perturbative barrier from the scale of
magnetostatic confinement (magnetic screening mass)
is indicated by the thick red line marked ``non-perturbative''.
The region below
it and up to $x=1$ is the regime of electrostatic QCD (EQCD),
while for $x\ge1$ the relevant effective theory is
given by non-static hard dense loops (HDL).
\label{fig:orders}}
\end{center}
\end{figure}

In Fig.~\ref{fig:orders} we display the structure of
the weak coupling expansion by showing how the magnitudes of its terms depend on the
power $x$ in the order-of-magnitude equality $T \sim g^x \mu$
(which of course removes the superconducting phase from the picture).
The $x$-axis in this figure corresponds to the ratio $\ln(T/\mu)/\ln(g)$
in the limit $g\to0$ and $T/\mu\to 0$.
For each value of $x$ we give the orders of the first several
terms in the expansion of the pressure in powers of $g$,
not counting separately terms with an extra factor of $\ln(g)$.
Full lines denote known contributions, while dashed and dash-dotted
lines correspond to the as-yet unknown ones.

At $x=0$, \textit{i.e.~}$T\sim \mu$ the
situation is still the same as with $\mu=0$: the weak coupling
expansion of the pressure is organized
in single powers (and logs) of $g$. The relevant effective
theory is given by dimensionally reduced electrostatic QCD (EQCD) which for the pressure has been worked out
up to but not including order $g^6$ which is where nonperturbative
physics from magnetostatic QCD (MQCD) starts to contribute.
This barrier is indicated by a thick red line in Fig.~\ref{fig:orders}.

Because $g$ is treated as an arbitrarily small parameter, everything in Fig.~\ref{fig:orders}
with the exception of the border at $x=0$ corresponds to the regime $T\ll\mu$,
namely $T\sim g^x\mu$ with $x>0$. As long as $x<1$, $T$ is parametrically
larger than the Debye mass $\sim g\mu$, and so dimensional reduction
should still be applicable.
However, each coefficient of the original series at $x=0$ now
has to be expanded in powers of $T/\mu\sim g^x$. The 2-loop
pressure contribution for example yields three different terms for $x>0$: one
is proportional to $\mu^4$ and thus is always of order $g^2$,
another --- proportional to $\mu^2 T^2$ --- gives the line $y=2+2x$
and the third term proportional to $T^4$ produces the line $y=2+4x$.
Starting with the plasmon term which is of order $g^3$ at $x=0$,
we obtain an infinite series of higher-order terms for $x>0$.
These arise from the expansion of the third power of the Debye mass parameter in powers of $T/\mu$, and, for
subsequent terms in the dimensional reduction result, also from the expansion of the special
functions $\aleph(n,z)$. Because both the Debye mass
and the $\aleph$ functions can be expanded
in even powers of $T/\mu$, the lines emanating from
their starting points at $x=0$ come with slopes differing
by two units. The terms proportional to $g^3$ and $g^5$
at $x=0$ involve a single overall power of $T$, so the lines
emanating from these have slopes 1, 3, 5, \ldots, whereas
the term proportional to $g^4$ (or $g^4 \ln\,g$) has
also $T$-independent parts and thus gives rise
to lines with slopes 0, 2, 4, \ldots.
In Eq.~(\ref{pDRsqrtg}) we have seen how this gives rise to a new series
in $g$ at $x=1/2$, and Fig.~\ref{fig:orders} illustrates
how the individual terms of order 2, 3, $7\over2$, 4, $9\over2$, \ldots
are produced from the various coefficients of the expansion at $x=0$.

Moving on to the border of applicability of the dimensional reduction
results, $x=1$, we see that all lines converge to points
corresponding to an expansion in even powers of $g$ (and
also involving $\ln\,g$). As noted before, for $x\ge 1$ the relevant effective
theory is the one
given by non-static hard dense loops. Their resummation
is necessary to obtain the classic Freedman-McLerran (FMcL) result
to order $g^4$ (again accompanied by a logarithmic term) at $T=0$
as well as the leading thermal corrections to the interaction
pressure. In a low-$T$ expansion
these $T$-dependent terms start with a contribution of order
$g^2 T^2 \mu^2 \ln(T/g\mu)$ and then involve fractional powers
$T^{8/3}$, $T^{10/3}$, $T^4\,\ln\, T$, $T^{14/3}$, \ldots$\;$such that the corresponding lines  in
Fig.~\ref{fig:orders} (labeled by the exponent of $T$)
meet at $x=1$ and effective order $g^4$.
At this point, the leading $T$-dependent contributions are of the
same order as the three-loop $T=0$ (FMcL) pressure contribution
and remain more important than the undetermined four-loop $T=0$ term even for
parametrically lower temperatures as long as $x<2$ (i.e.\ $T\gg g^2 \mu$).
For the entropy and specific heat, for which the zero-temperature contribution to
the pressure drops out, these $T$-dependent terms represent
the leading interaction contributions down to arbitrarily low temperatures.
The $T\,\ln\, T$ behavior of the entropy (as well as of the specific heat) is
characterized by ``anomalous'' non-Fermi-liquid behavior, caused by the only
weakly (dynamically) screened quasi-static magnetic interactions
with an effective frequency-dependent screening mass, displayed in Eq.~(\ref{mmdyn}).

As suggested by Fig.~\ref{fig:orders} and
shown in detail in the previous section, the HDL-resummed
thermal pressure contributions responsible for the
non-Fermi-liquid behavior at $T\ll g\mu$ match smoothly
to the perturbative effects at $T\gg g\mu$ described by
EQCD. As the temperature is increased, electrostatic
screening replaces dynamical magnetic screening
as the dominant collective phenomenon also in the $T$-dependent
contributions.
For $T$ parametrically larger than $g\mu$ (\textit{i.e.~}$x<1$)
the resummation of HDL self energies needs to be trivially extended
to HTL self energies to avoid accuracy loss.
When added to the zero temperature ${\mathcal O}(g^4)$ result,
this gives an expression that gives the pressure
for all temperatures and chemical potentials up to
an error of order $g^{{\rm min}(4+2x,6)}$ (or $g^{4+x}$ throughout
in the case of the entropy, for which the unknown four-loop T=0 pressure
drops out).

From the ``flow'' of the various perturbative
contributions as a function of $x$ in Fig.~\ref{fig:orders}, one notices that a single expression aiming to be
valid both for $x>1$ and $x<1$ needs to keep track of
contributions which are perhaps higher-order and irrelevant
in some region but essential in another.
The novel approach we have presented here does so
by resumming the complete one-loop gluon self-energy in all
IR sensitive graphs while treating the infrared-safe
2GI diagrams perturbatively. To the extent that we have worked
it out, this procedure covers both $x>1$ and $x<1$ with an error of order
$g^{{\rm min}(5+x,6)}$ which improves over the HTL/HDL result in the
region $x<1$ by including the contributions of all relevant three-loop
graphs. A drawback compared to the HTL/HDL resummation schemes is however
that the resummation of the complete gluon self-energy
leads to gauge-dependent higher-order contributions whose unphysical
nature is highlighted by the appearance of spacelike poles in the
logarithmic resummation integrand with momenta $\sim g^2T$ and also
of an unphysical damping constant (with an incorrect sign) $\propto g^2 T$.
For our expression for the pressure, the effect of these problems is, however,
only of the order of the nonperturbative MQCD contributions, \textit{i.e.}~$g^6$,
so it has not hindered us from confirming and thus validating the
results obtained through dimensional reduction or the HTL/HDL approach.

\section{Conclusions and outlook}\label{sec:concl}

In this paper, we have constructed a novel resummation scheme designed to reproduce the weak coupling expansion of the QCD pressure up to order $g^4$
on the entire $\mu$-$T$ plane. We have used it to provide an independent check of practically all existing perturbative results. In particular,
we have performed
the first explicit test on the validity of dimensional reduction for values of $\mu/T$ far beyond the capability of present-day lattice
techniques, thus verifying that dimensionally reduced effective theories provide a solid description of the perturbative physics up to in principle arbitrarily large values
of $\mu/T$ as long as $\pi T>m_\rmi{D}$. At temperatures parametrically smaller than the chemical potential, we have on the other hand reproduced numerically all the results of the
HTL/HDL resummation schemes, verifying their validity and highlighting the smooth transition taking place in the perturbative expansion of the pressure
as one moves from the region of dimensional reduction towards the zero-temperature limit.

Based on our numerical results from Section \ref{sec:numres},
the dimensional reduction result for the QCD pressure appears to be provide a remarkably good
approximation for this quantity down to the point where
the $T$-dependent contribution to the interaction pressure, $\delta p$,
ceases to be negative (cf.\ Figs.\ \ref{fig:p01}ff)
which happens at $T\approx 0.2 m_\rmi{D}$. Since the
dimensional reduction result to order $g^6\ln\,g$ combined
with optimized choices of the renormalization scale
has turned out to agree rather well with
lattice results, both at zero chemical potential
\cite{klry,Blaizot:2003iq} and
for $\mu\sim T$ \cite{avpres,Ipp:2003yz}, our present findings in fact suggest a remarkably wide
practical range of applicability for the dimensional reduction method and results.

Progressing down on the temperature axis to $T\lesssim 0.2 m_\rmi{D}$, one eventually has to switch to the nonstatic resummation schemes provided either by
our new approach or by the calculationally much simpler HTL resummation of Eq.~(\ref{PHTLtot}).
At such low temperatures, the pressure can --- up to but not including order $g^6$ --- be approximated by the Freedman-McLerran result plus positive contributions from the
Stefan-Boltzmann terms as well as the interaction pressure $\delta p$. The latter of these is the source of the non-Fermi-liquid behavior of the entropy and specific heat.

While we believe to have thoroughly clarified the nature of perturbative
expansions of the pressure in different regimes of the $\mu$-$T$ plane,
our new approach is, as of today, yet to produce results for the pressure beyond what has already been achieved through either dimensional reduction at
$x<1$, the HTL/HDL resummation schemes at $x\geq 1$ or the Freedman-McLerran result at $T=0$. Its present relative error of order
$g^{{\rm min}(5+x,6)}$ can in principle be reduced through the inclusion of the two-loop gluon polarization tensor into the resummation of the
ring diagrams and in addition by taking the contributions of non-static modes into account in the multiple sums of Fig.~2.b-d. For the pressure,
this would bring the accuracy of our new approach up to the one currently achieved by dimensional reduction calculations (excluding the already known
${\mathcal O}(g^6\ln\,g \,T^2(T^2+\mu^2))$ term), so that the error, up to logarithms, would be uniformly (for all values of $x$) of order $g^6 \ln\,g$,
corresponding to the line marked ``4-loop $T=0$ pressure'' in Fig.~\ref{fig:orders}.
This would then unify all existing perturbative results for the pressure of QCD, while for the entropy it would moreover lead to genuinely
new results. Apart from increasing \textit{e.g.}~the accuracy of the entropy result at $x=1/2$ to order $g^{13/2}$
(green open dots in Fig.~\ref{fig:orders})\footnote{The highest purely perturbatively calculable order at $x=1/2$
is $g^{15/2}$ which would require a calculation of the contributions of order $g^6\mu^2 T^2$ and $g^7\mu^3T$ for the pressure.},
it would push the error in the $T$-dependent part of the pressure up to the line denoted in Fig.~\ref{fig:orders} by ``4-loop T contribution''
and thus, for $x>1$, include the so far unknown order $g^4\mu^2 T$ corrections to the non-Fermi-liquid terms in the entropy and the specific heat.
Considering the difficulties caused by the gauge-dependent parts of the gluon self-energy, it seems
that such an extension should probably aim at keeping only gauge-independent contributions such as HTL self energies
in the ring diagrams and treating corrections to those self-energies in a perturbative manner.
Work towards this goal is currently in progress.

\acknowledgments

We are grateful to Mikko Laine, York Schr\"oder,
and Larry Yaffe for their helpful
comments and suggestions and to Dirk Rischke for discussions
on color superconductivity. This
work has been partially supported by the Austrian Science Foundation
FWF, project no.\ P16387-N08 and the Academy of Finland, project no. 109720.


\appendix

\section{Dimensional reduction result at finite $T$ and $\mu$}

To order $g^5$ and simplified by assuming equal chemical potentials,
the result of dimensional reduction for the QCD pressure at finite $T$ and $\mu$,
obtained in Ref.\ \cite{avpres}, reads
\ba\label{dimredpr5}
 \frac{p_\rmi{DR}(T,\mu)}{T^4} & = &
 \fr{\pi^2}{45}\bigg\{d_A+\bigg(\fr{7}{4} + 30\mubar^2 +
60\mubar^4\bigg)d_F\bigg\}\nn
&-&g^2\fr{d_A}{144}\bigg\{C_A +
\fr{T_F}{2}\(1+12\mubar^2\)\(5+12\mubar^2\)\!\bigg\}\nn
&+& \frac{g^3}{(4\pi)}\frac{d_A}{3}
\Big\{\fr{1}{3}\(C_A+T_F\(1+12\mubar^2\)\)\Big\}^{3/2} \nn
 & + &
 \frac{g^4}{(4\pi)^2}
 \fr{d_A}{144}\bigg\{48C_A\(C_A+T_F\(1+12\mubar^2\)\)\ln\Big[\!\(C_A+T_F\(1+12\mubar^2\)\)^{1/2}g\Big]\nn
&-&C_A^2\bigg(\fr{22}{3}\ln\fr{\bar{\Lambda}}{4\pi T} + \fr{64}{5}
+24\,\ln\,12\pi^2- 4\gamma +\fr{38}{3}\fr{\zeta'(-3)}{\zeta(-3)} -
\fr{148}{3}\fr{\zeta'(-1)}{\zeta(-1)}\bigg) \nn
&-& C_A T_F\bigg(\bigg(\fr{47}{3}+264\mubar^2+528\mubar^4\bigg)\ln\fr{\bar{\Lambda}}{4\pi
T} +\fr{1759}{60} + 8\gamma +24(1+12\mubar^2)\,\ln\,(12\pi^2)\nn
&+& 2\(161-48\gamma\)\mubar^2 - 644\mubar^4
- \fr{268}{15}\fr{\zeta '(-3)}{\zeta(-3)} -
\fr{4}{3}\(11+156\mubar^2\)\fr{\zeta'(-1)}{\zeta(-1)} \nn
&-& 24\Big[52\,\aleph(3,z)
+
144\imathb\mubar\,\aleph(2,z)+\(5-92\mubar^2\)\aleph(1,z)+4\imathb\mubar\,\aleph(0,z)\Big]\bigg)
\nn
&+& C_F T_F
\bigg(\fr{3}{4}\(1+4\mubar^2\)\(35+332\mubar^2\)-24\(1-4\mubar^2\)\fr{\zeta'(-1)
}{\zeta(-1)}
\nn
&-& 144\Big[12\imathb\mubar\,\aleph(2,z)-2\(1+8\mubar^2\)\aleph(1,z)
-\imathb\mubar\(1+4\mubar^2\)\aleph(0,z)\Big]\bigg) \nn
&+& T_F^2 \bigg(\fr{4}{3}\(1+12\mubar^2\)\(5 +
12\mubar^2\)\ln\fr{\bar{\Lambda}}{4\pi T}
+ \fr{1}{3}+4\gamma + 8\(7+12\gamma\)\mubar^2\nn
&+& 16(43+36\gamma)\mubar^4
- \fr{64}{15}\fr{\zeta'(-3)}{\zeta(-3)}
- \fr{32}{3}\(1+12\mubar^2\)\fr{\zeta'(-1)}{\zeta(-1)} \nn
&-& 96\Big[8\,\aleph(3,z) + 12\imathb\mubar\,\aleph(2,z) -
2\(1+8\mubar^2\)\aleph(1,z)\nn
&-&
\imathb\mubar(1+12\mubar^2)\,\aleph(0,z)+3\aleph(3,2z)+12\imathb\mubar\aleph(2,2z)-12\mubar^2\aleph(1,2z)\Big]\bigg)\bigg\}
\nonumber\\
& + &
\frac{g^5}{(4\pi)^3}d_A\sqrt{\fr{1}{3}\big(C_A+T_F\(1+12\mubar^2\)\big)}
\bigg\{-C_FT_F(1+12\mubar^2)\nn
&+& C_A^2\bigg(\fr{11}{9}\ln\fr{\bar{\Lambda}}{4\pi T} - \fr{247}{72}
+\fr{11}{6}\,\ln\,2+ \fr{11}{9}\gamma-\fr{\pi^2}{6}\bigg)\\
&+&C_AT_F\bigg(\bigg(\fr{7}{9}+\fr{44}{3}\mubar^2\bigg)\ln\fr{\bar{\Lambda}}{4\pi
T} +\fr{1}{2} + \fr{11}{9}\gamma
+\fr{22}{3}\(1+2\gamma\)\mubar^2+\fr{2}{9}\aleph(z)\bigg) \nn
&-&T_F^2\bigg(\bigg(\fr{4}{9}+\fr{16}{3}\mubar^2\bigg)\ln\fr{\bar{\Lambda}}{4\pi
T} -\fr{2}{9}
-\fr{8}{3}\mubar^2-\fr{2}{9}\(1+12\mubar^2\)\aleph(z)\bigg)\bigg\}  +
{\cal O}(g^6\ln\,g).\nonumber
\ea

\section{The analytic values of various parts of the pressure}
\label{app:anl}

In this Appendix, we collect some calculational details on
the different pieces of the analytical part of the
pressure, $p_\rmi{anl}$, as defined in (\ref{res1}).
The sum of the 2GI diagrams was already given
in Eq.~\nr{2pi} and the piece $p_\rmi{b}$ in Eq.~(\ref{pbres}).

\subsection{The Vac-Vac and Vac-Mat diagrams}\label{sec:anlI}

The evaluation of the two special diagrams of Fig.~2, dubbed Vac-Vac and Vac-Mat based on their self energy insertions, is relatively straightforward.
Inserting the form of $\Pi_{\mu\nu}^{ab}(P)\!\!\mid_{\rmi{vac}}$ from Eq.~(\ref{polarvac}) into the Feynman gauge expressions of the graphs, contracting
all Lorentz and color indices and taking into account the symmetry factors $1/4$ and $1/2$, respectively, we readily obtain
\ba\label{ivv}
p_\rmi{VV}&=&\fr{1}{4}(d-1)d_A\tilde{A}^2g^4\Lambda^{4\e}\I_{2\e}\\
&=&\fr{d_A(5C_A-4T_F)^2}{4320(4\pi)^2}g^4T^4\bigg(\fr{1}{\e}+6\,\ln\fr{\bar{\Lambda}}{4\pi T}+\fr{32}{3}+6\fr{\zeta'(-3)}{\zeta(-3)}+\fr{4C_A}{5C_A-4T_F}\bigg),
\nn
p_\rmi{VM}&=&-2p_\rmi{VV}-\fr{1}{2}d_A\tilde{A}g^4\Lambda^{2\e}\bigg\{\(C_A(d-2)^2\I_1-4T_F(d-2)\It_1\)\I_{1+\e}\\
&-&
(5-3\e)C_A\sumint_{PQ}\fr{1}{(P^2)^{\e}Q^2(P+Q)^2}+2(d-2)T_F\sumint_{P\{Q\}}\fr{1}{(P^2)^{\e}Q^2(P+Q)^2}\bigg\},\nonumber
\ea
where we in the latter case have two two-loop sum-integrals to evaluate. These can be taken care of using methods developed in Refs.~\cite{az,avpres},
which leads to the final result
\ba
p_\rmi{VM}&=&-2p_\rmi{VV}+\fr{d_Ag^4T^4}{4320(4\pi)^2}\bigg\{\fr{(5C_A-4T_F)(10C_A-(29+360\mubar^2+720\mubar^4)T_F)}{\e}\nn
&+&100C_A^2\(3\,\ln\fr{\bar{\Lambda}}{4\pi T}+\fr{463}{60}-\fr{\zeta'(-1)}{\zeta(-1)}+4\fr{\zeta'(-3)}{\zeta(-3)}\)\nn
&+&30C_AT_F\bigg(-\(37+360\mubar^2+720\mubar^4\)\ln\fr{\bar{\Lambda}}{4\pi T}-\fr{8971}{180}-\fr{32}{3}\za-\fr{40}{3}\zb \nn
&-&2\(167+80\za\)\mubar^2-668\mubar^4-240(1+4\mubar^2)\aleph(1,z)-480\aleph(3,z)\bigg)\nn
&+&24T_F^2\bigg(\(29+360\mubar^2+720\mubar^4\)\ln\fr{\bar{\Lambda}}{4\pi T}+\fr{1003}{36}+\fr{40}{3}\za+\fr{8}{3}\zb \nn
&+&10\(31+16\za\)\mubar^2+620\mubar^4+240(1+4\mubar^2)\aleph(1,z)+480\aleph(3,z)\bigg)\bigg\}. \label{ivm}
\ea

\subsection{The ordinary ring sum}\label{app:ringsum}

The UV subtraction term $p_\rmi{ring}^\rmi{UV}$ of the ordinary ring sum, defined in Eq.~(\ref{pUV}),
is straightforward to evaluate and produces the result
\ba
p_\rmi{ring}^\rmi{UV}&=&(3+4\e)d_AC_A^2g^4\(\I_1\)^2\I_2\nn
&=&\fr{d_AC_A^2}{48(4\pi)^2}g^4T^4\bigg(\fr{1}{\e}+6\,\ln\fr{\bar{\Lambda}}{4\pi T}+\fr{16}{3}+
2\gamma+4\za\bigg)+\mathcal{O}(\e). \label{puv}
\ea
With the IR regulating term $p_\rmi{ring}^\rmi{IR}$,
we find it most convenient to first transform the Matsubara sum into a contour integral in the standard way. After the
term proportional to a bosonic distribution function is analytically continued to Minkowski space (see Appendix D), this leads to the result
\ba
p_\rmi{ring}^\rmi{IR} &=&\fr{d_AC_A^2g^4T^4}{48}\Bigg(-{\cal I}_{2}^0+\Lambda^{2\e}\int\fr{d^{3-2\e}p}{(2\pi)^{3-2\e}}\bigg\{\fr{1}{2\pi i}\int_{-i\infty}^{i\infty}dp_0
\fr{1}{(p_0^2-p^2-m^2)^2}\nn
&-&\fr{1}{\pi}\int_{0}^{\infty}dp_0 n_b(p_0){\rm Im}\Big[
\fr{1}{((p_0+i\e)^2-p^2-m^2)^2}\Big]\bigg\}\Bigg).
\label{f}
\ea
The sum of the first two terms is evaluated with standard methods and gives
\ba
-{\cal I}_{2}^0+\fr{\Lambda^{2\e}}{2\pi i}\int\fr{d^{3-2\e}p}{(2\pi)^{3-2\e}}\int_{-i\infty}^{i\infty}dp_0
\fr{1}{(p_0^2-p^2-m^2)^2}
&=&-\fr{2}{(4\pi)^2}\Big(\ln\fr{m}{4\pi T}+\gamma\Big),\label{euclmass1}
\ea
while the third term (where one can set $\e=0$) is left in its present form until Appendix D where it is seen to explicitly cancel with certain parts of $p_\rmi{ring}^\rmi{finite}$.
Modifications to $p_\rmi{ring}^\rmi{IR}$ induced by a magnetic mass are discussed after Eq.~(\ref{eq:pE3}).



\subsection{Small  and large $\mu /T$ limits of $p_\rmi{anl}$}

For convenience, we provide here formulae for various limits of the function $p_\rmi{anl}$
of Eq.~(\ref{panl}), derived using the results of Appendix D of
Ref.~\cite{avpres} for the behavior of the $\aleph$ functions. First, in the limit
$\mu\rightarrow 0$ we obtain through an expansion in $\mubar\equiv \fr{\mu}{2\pi T}$
\ba
p_\rmi{anl}&\xrightarrow[\mu\rightarrow 0]{}&\pi^2d_AT^4\Bigg\{\fr{1}{45}
\bigg\{1+\fr{d_F}{d_A}\(\fr{7}{4}+30\mubar^2\)\bigg\} - \fr{g^2}{9(4\pi)^2}\bigg\{C_A +
\fr{T_F}{2}(1+72\mubar^2)\bigg\}\nn
&+&\fr{g^4}{27(4\pi)^4}\bigg\{-C_A^2\bigg(22\,\ln\fr{\bar{\Lambda}}{4\pi T}+63-18\gamma
+110\za-70\zb\bigg)\nn
&-&C_AT_F\bigg(\(47+792\mubar^2\)\ln\fr{\bar{\Lambda}}{4\pi T}+
\fr{2391}{20}+70\za-23\zb-\fr{297}{5}\ln\,2 \nn
&+&6\(235+352\,\ln\,2+132\gamma\)\mubar^2\bigg)\nn
&+&T_F^2\bigg(4(5+72\mubar^2)\ln\fr{\bar{\Lambda}}{4\pi T}+\fr{99}{5}-\fr{108}{5}\ln\,2+40\za-20\zb\nn
&+&24\(11+32\,\ln\,2+12\gamma\)\mubar^2\bigg)\nn
&+&\fr{9}{4}C_FT_F\Big(35-32\,\ln\,2+8(35+16\,\ln\,2)\mubar^2\Big)\bigg\}+
\mathcal{O}(\mubar^4)\Bigg\},
\label{mutozero}
\ea
which contains \textit{e.g.}~the $C_FT_F$ dependent part of the linear quark number susceptibilities
found in Ref.~\cite{avsusc}. Taking the opposite limit,
$T\rightarrow 0$, we on the other hand get
\ba
p_\rmi{anl}&\xrightarrow[T\rightarrow 0]{}&\fr{d_A\mu^4}{(4\pi)^2}\Bigg\{4\(\fr{d_F}{3d_A}-2T_F\fr{g^2}{(4\pi)^2}\)\(1+\fr{1}{2\mubar^2}\)\nn
&-&\fr{8}{3}\fr{g^4}{(4\pi)^4}\bigg\{11C_AT_F\bigg(2\,\ln\fr{\bar{\Lambda}}{2\mu}+\fr{142}{33} \nn
&+&\bigg[\ln\fr{\bar{\Lambda}}{4\pi T}-\fr{1}{3}\ln\mubar+\fr{395}{132}-
\fr{48}{11}\ln\,2+\fr{2}{3}\za\bigg]\mubar^{-2}\bigg)\label{ttozero}\\
&-&4T_F^2\bigg(2\,\ln\fr{\bar{\Lambda}}{2\mu}+\fr{11}{3} +
\bigg[\ln\fr{\bar{\Lambda}}{4\pi T}-\fr{1}{3}\ln\mubar+\fr{19}{12}+\fr{2}{3}\za\bigg]\mubar^{-2}\bigg)\nn
&-&C_FT_F\bigg(\fr{51}{2}+
\bigg[4\,\ln\mubar+\fr{35}{4}+4\za\bigg]\mubar^{-2}\bigg)\bigg\}+
\mathcal{O}(\mubar^{-4}\ln\,\mubar)\Bigg\}.\nonumber
\ea
This yields an entirely finite expression at $T=0$, where the scale
dependence of the result naturally coincides with that of Eq.~(\ref{pFMcL}).

\section{Properties of the one-loop gluon polarization tensor}
\label{app:Pi}

For reference, we note that the one-loop gluon polarization tensor in the Feynman gauge can be written in the form
\ba\label{polar}
\Pi_{\mu\nu}^{ab}(P)&=& g^2 \delta^{ab}\bigg\{C_A\bigg((d-2)\I_1\delta_{\mu\nu}+2\(P_{\mu}P_{\nu}-P^2\delta_{\mu\nu}\)\Pi\!\(P\)\nn &-&
\fr{d-2}{2}\sumint_{Q}\fr{(2Q-P)_{\mu}(2Q-P)_{\nu}}{Q^2(Q-P)^2}\bigg)\\
&-&2T_F\bigg(2\It_1\delta_{\mu\nu}+\(P_{\mu}P_{\nu}-P^2\delta_{\mu\nu}\)\Pi_\rmi{f} \!\(P\)-
\sumint_{\{Q\}}\fr{(2Q-P)_{\mu}(2Q-P)_{\nu}}{Q^2(Q-P)^2}\bigg)\bigg\}, \nonumber
\ea
where we have, as usual, defined
\ba
\Pi(P)&\equiv&\sumint_Q\fr{1}{Q^2(Q-P)^2}, \\
\Pi_\rmi{f}(P)&\equiv&\sumint_{\{Q\}}\fr{1}{Q^2(Q-P)^2}.
\ea
The vacuum ($T,\mu\rightarrow 0$) limit of the above formula is uniquely determined by replacing the sum-integrals by
$4-2\e$-dimensional integrals which produces
\ba
\Pi_{\mu\nu}^{ab}(P)\mid_{\rmi{vac}}&=&\tilde{A}g^2\delta^{ab}\(\fr{\Lambda^2}{P^2}\)^{\!\!\e}\(P_{\mu}P_{\nu}-
P^2\delta_{\mu\nu}\) \label{polarvac}
\ea
with
\ba
\tilde{A}&\equiv&\fr{((5/2+\e)C_A-2T_F)}{24\pi^2}\(\fr{1}{\e}-\gamma+\ln (4\pi)+\fr{5}{3}+\mathcal{O}(\e)\).
\ea

As all our analytic formulae, the above expressions have been given in Euclidean space. For the numerical evaluation of $p_\rmi{ring}^\rmi{safe}$ it is, however, convenient to also
reproduce the known formulae for the transverse and longitudinal pieces of the vacuum-subtracted self energy in Minkowski space. Following Ref.~\cite{Weldon:1982aq}, we obtain
\begin{eqnarray}
\Pi_\rmi{L}(q_{0},q) & = & -g^{2}\frac{Q^{2}}{q^{2}}\left(2T_F H_{f}+C_A H_{b}\right),\label{PiL}\\
\Pi_\rmi{T}(q_{0},q) & = & \frac{g^{2}}{2}\left(\frac{Q^2}{q^{2}}\left(2 T_F H_{f}+C_A H_{b}\right)+\left(2 T_F G_{f}+C_A G_{b}\right)\right),\label{PiT}
\end{eqnarray}
where the different functions read
\begin{eqnarray}
G_{f} & = & \frac{1}{2\pi^{2}}\int_{0}^{\infty}dk\, n_{f}(k)\left(4k-\frac{q^{2}-q_{0}^{2}}{2q}L_{1}\right),\label{Gf}\\
H_{f} & = & \frac{1}{2\pi^{2}}\int_{0}^{\infty}dk\, n_{f}(k)\left(2k-\frac{q^{2}-q_{0}^{2}-4k^{2}}{4q}L_{1}-q_{0}kL_{2}\right),\label{Hf}\\
G_{b} & = & \frac{1}{2\pi^{2}}\int_{0}^{\infty}dk\, n_{b}(k)\left(4k-\frac{5}{4}\frac{q^{2}-q_{0}^{2}}{q}L_{1}\right),\label{Gb}\\
H_{b} & = & \frac{1}{2\pi^{2}}\int_{0}^{\infty}dk\, n_{b}(k)\left(2k-\frac{2q^{2}-q_{0}^{2}-4k^{2}}{4q}L_{1}-q_{0}kL_{2}\right),\label{Hb}
\end{eqnarray}
with
\begin{eqnarray}
L_{1} & = & \log\left(\frac{2k+q-q_{0}}{2k-q-q_{0}}\right)-\log\left(\frac{2k-q+q_{0}}{2k+q+q_{0}}\right),\label{L1}\\
L_{2} & = & \log\left(\frac{2k+q-q_{0}}{2k-q-q_{0}}\right)-2\log\left(\frac{-q+q_{0}}{q+q_{0}}\right)+\log\left(\frac{2k-q+q_{0}}{2k+q+q_{0}}\right)\label{L2}.
\end{eqnarray}
These expressions are valid for all complex $q_{0}$ in a rotation
from Euclidean space $q_{0}=i\omega$ to Minkowski space $q_{0}=\omega+i\epsilon$
with $\epsilon>0$. For the analytic continuation into the region
with $\epsilon<0$, see Ref.~\cite{Blaizot:2005fd}.

The corresponding HTL and HDL expressions for the above functions can be extracted by demanding
that the external momenta $q_{0}$ and $q$ be small compared to the
temperature $T$ and/or the chemical potential $\mu$. This amounts to expanding the integrand in inverse powers of $k$ to leading order, since
the main contribution is expected to come from large loop momenta $k$. The relevant physical scale is given by the Debye
mass $m_{\rmi D}$. In this limit, the equations (\ref{Gf}) to (\ref{Hb}) reduce to
\begin{eqnarray}
G_{f}^\rmi{HTL} & = &
\frac{T^{2}}{6}+\frac{\mu^{2}}{2\pi^{2}},\label{GfHTL}\\
H_{f}^\rmi{HTL} & = &
\left(\frac{T^{2}}{6}+\frac{\mu^{2}}{2\pi^{2}}\right)\left\{
1+\frac{q_{0}}{2q}\log\left(\frac{-q+q_{0}}{q+q_{0}}\right)\right\}
,\label{HfHTL}\\
G_{b}^\rmi{HTL} & = & \frac{T^{2}}{3},\label{GbHTL}\\
H_{b}^\rmi{HTL} & = & \frac{T^{2}}{3}\left\{
1+\frac{q_{0}}{2q}\log\left(\frac{-q+q_{0}}{q+q_{0}}\right)\right\}
.\label{HbHTL}\end{eqnarray}
The HDL expressions are obtained by setting $T=0$ here which makes both $G_{b}^\rmi{HDL}$ and $H_{b}^\rmi{HDL}$ vanish.

\section{Numerical evaluation of $p_\rmi{ring}^\rmi{finite}$}\label{app:num}

The numerical evaluation of the sum-integrals introduced in Sec.~III is performed by converting the sums over Matsubara frequencies into contour integrals where the integrand is multiplied by
$\cot(\omega_n/2T)$ and the contour encircles the poles of this function (which lie on the real axis at $\omega_n=2\pi nT$ and have the residue $2T$). We then change the integration
variable from $q_0 = i\omega_n$, with real $\omega_n$, to $q_0 + i\epsilon$, with real $q_0$ and $\epsilon > 0$, deforming the integration contour in the usual way (for more details on this,
see \textit{e.g.~}Ref.~\cite{kap}). Through this procedure, we obtain the generic result
\begin{eqnarray}
T\!\!\!\sum_{\omega_n=2\pi nT}^{n\in Z}f(i\omega_n) & = & \int_{0}^{\infty}\frac{dq_0}{\pi}\coth\left(\frac{q_0}{2T}\right)
\frac{-i}{2}\left\{ f(q_0+i\epsilon)-f^{*}(q_0+i\epsilon)\right\} \nonumber \\
 & = & \int_{0}^{\infty}\frac{dq_0}{\pi}\left(1+2n_b\right){\rm Im}\, f(q_0+i\epsilon)\label{eq:Matsubaraconversion}
\end{eqnarray}
where we have used the identity $\coth\frac{q_0}{2T}=1+2n_b$.
This is valid for any function $f$ satisfying $f(q_0+i\epsilon)=f^{*}(q_0-i\epsilon)=f(-q_0-i\epsilon)=f^{*}(-q_0+i\epsilon)$,
provided that the great arc contribution to the integral vanishes and there are no other poles or branch cuts between the Euclidean and Minkowskian axes.

After these standard manipulations, we split the above integral into the UV safe ``$n_{b}$ part'' which we calculate
in Minkowski space, \textit{i.e.~}along the real $q_0$ axis, and the ``non-$n_{b}$ part'' that we
rotate back to Euclidean space\footnote{The quantities evaluated in Minkowskian metric $(+,-,-,-)$ (implying $Q^2=q_0^2 - q^2$) are denoted by an upper index
M, while Euclidean pieces, where $q_0$ is analytically continued back to Euclidean space ($q_0 \rightarrow -i q_0 \equiv\omega$), are indicated by the use of the ``Matsubara variable'' $\omega$
and the upper index E. We have dropped the index $n$ from $\omega_n$ as it now has become a continuous variable.
}
before integration through
\begin{eqnarray}
\int_{0}^{\infty}\frac{dq_0}{\pi}{\rm Im}\, f(q_0+i\epsilon) & \rightarrow & \int_{0}^{\infty}\frac{d\omega}{\pi}{\rm Re}\, f(i\omega).
\end{eqnarray}
In this way, we take full advantage of the UV cutoff properties of $n_b$ in Minkowski space while the possibly UV problematic non-$n_b$ parts are
treated respecting the Euclidean symmetry. Further discussion on the reasons for this separation can be found from Refs.~\cite{ippreb,moore}.

Specializing to the evaluation of the function $p_\rmi{ring}^\rmi{finite}$, there is a further subtlety in the non-$n_b$ part that we have to take into account. As discussed in
Sec.~\ref{sec:numiss},
there is an unphysical pole in the integrand of Eq.~(\ref{eq:pringfinite}) which implies that we cannot deform our integration path from the Euclidean to the Minkowskian axis quite as
suggested by Eq.~(\ref{eq:Matsubaraconversion}). Instead, we have to try to avoid numerically dangerous singularities on the complex $q_0$ plane by encircling them at a safe distance. This
aspect of our calculation is discussed in more detail below in Appendix \ref{sed:contourdeformation}.

After all the separations, we can write the function $p_\rmi{ring}^\rmi{finite}$ as the sum of several individual contributions according to
\begin{equation}
p_\rmi{ring}^\rmi{finite}=p_\rmi{1}^\rmi{E}+p_\rmi{2}^\rmi{E}+p_\rmi{L}^\rmi{M}+p_\rmi{LD}^\rmi{M}+p_\rmi{T}^\rmi{M}+p_\rmi{TD}^\rmi{M}+p_\rmi{$m$}^\rmi{M}.\label{eq:pfinringv2}
\end{equation}
Here, $p_\rmi{1}^\rmi{E}$ and $p_\rmi{2}^\rmi{E}$ denote two distinct parts of the Euclidean integral that are separated by a
momentum cut-off $\Lambda$. The pieces $p_\rmi{L}^\rmi{M}$ and $p_\rmi{T}^\rmi{M}$ on the other hand stand for longitudinal and transverse Minkowskian
contributions to $p_\rmi{ring}^\rmi{finite}$ integrated along specific segments of the real $q_0$-axis and $p_\rmi{LD}^\rmi{M}$ and $p_\rmi{TD}^\rmi{M}$ for
the corresponding quantities integrated along deformed paths (see Fig.~(\ref{contdef})). In addition, $m$ refers to corrections due to the IR regulator mass introduced in Sec.~III C
that can be explicitly extracted from the rest. Each of these terms will be discussed in detail in the following.

\subsection{Euclidean space contributions}

For practical numerical reasons, we divide the Euclidean space ``non-$n_b$'' contribution to Eq.~(\ref{eq:pfinringv2}) into two parts $p_\rmi{1}^\rmi{E}+p_\rmi{2}^\rmi{E}$ by introducing
a four-momentum
cut-off $\Lambda$ and having the first piece correspond to the contribution of momenta with $|Q|<\Lambda$. For this, we obtain by first performing the integration over an Euclidean four-sphere
\begin{equation}
p_\rmi{1}^\rmi{E}=-\frac{d_A}{2\pi^{3}}\int_{0}^{\Lambda}dQ\,Q^{3}\int_{0}^{\pi/2}d\alpha\,\sin^{2}\alpha\,\re(p_\rmi{int}^\rmi{E}(\omega,q)),\label{eq:pE}
\end{equation}
with $\omega=-i q_0=Q\cos\alpha$ and $q=Q\sin\alpha$.
The integrand $p_\rmi{int}^\rmi{E}(\omega,q)$ is given by
\ba
p_\rmi{int}^\rmi{E}(\omega,q) & = & \log\left[1+\frac{\Pi_\rmi{T}+m_\rmi{mag}^{2}}{q^{2}+\omega^{2}}\right]-\frac{\Pi_\rmi{T}+m_\rmi{mag}^{2}}{q^{2}+\omega^{2}}
 +  \frac{\left(C_A g^{2}T^{2}-6m_\rmi{mag}^{2}\right)^{2}}{72\left(q^{2}+\omega^{2}+m^{2}\right)^{2}}\nn
 & + & \frac{1}{2}\left(\log\left[1+\frac{\Pi_\rmi{L}}{q^{2}+\omega^{2}}\right]-\frac{\Pi_\rmi{L}}{q^{2}+\omega^{2}}
  + \frac{C_A^{2}g^{4}T^{4}}{72\left(q^{2}+\omega^{2}+m^{2}\right)^{2}}\right)\label{eq:pE3}
\ea
where the magnetic mass $\mmag$, introduced in Eq.~(\ref{mmagf}), is needed in the transverse part
in order to prevent a negative argument of the logarithm as $\omega^{2}+q^{2}\rightarrow0$. This provides a cutoff for IR divergences at the scale
$\mmag=c_f \frac{3}{32}g^{2}C_A T$, with $c_f \geq1$ being a ``magnetic factor'' that we can vary to verify
that the dependence of the result on it is beyond ${\mathcal O}(g^4)$.

In addition to regulating IR divergencies, the inclusion of the magnetic mass alters the UV behavior of the first term of the above integrand which is why we
included it also in the coefficient of the UV regulating last term of the first line of Eq.~(\ref{eq:pE3}).
In order to keep the final result independent of the arbitrary mass parameter $m$, we have to take the magnetic mass $m_\rmi{mag}$ into account also in the
calculation of the function $p_\rmi{ring}^\rmi{IR}$ of Eqs.~(\ref{pring37}) and (\ref{f}), which amounts to replacing $C_A^2$ in those equations as well as later in Eqs.~(\ref{contrib_Eucl}) and (\ref{eq:pmM}) by $C_A^{2}-8g^{2}C_A\mmagb^{2}+24g^{4}\mmagb^{4}$, as shown below in Eq.~(\ref{eq:phighzero}).

\subsubsection{High momentum expansion in Euclidean space}

To obtain the high-momentum Euclidean contribution $p_\rmi{2}^\rmi{E}$ to Eq.~(\ref{eq:pfinringv2}) in the most effective way, we first integrate over $\alpha$, then expand $\Pi_\rmi{T}$ and
$\Pi_\rmi{L}$ in the limit of large momenta (which is possible analytically order by order in $Q/T$) and finally perform the integration over $Q$. This produces a series of the general form
\begin{equation}
p_\rmi{2}^\rmi{E}=p_\rmi{high,0}^\rmi{E}+p_\rmi{high,2}^\rmi{E}+p_\rmi{high,4}^\rmi{E}+...
\end{equation}
where the lowest orders are given by
\ba
p_\rmi{high,0}^\rmi{E}  &=&  -\frac{d_A}{768\pi^{2}}g^{4}T^{4}\left(C_A^{2}-8g^{2}C_A \mmagb^{2}+24g^{4}\mmagb^{4}\right)
\left(\log\frac{\Lambda^{2}}{m^{2}+\Lambda^{2}}
-\frac{m^{2}}{m^{2}+\Lambda^{2}}\right),\label{eq:phighzero}\\
p_\rmi{high,2}^\rmi{E} & = & -\frac{d_A}{3456\pi^{2}\Lambda^{2}}g^{6}T^{6}\left\{ -\frac{C_A^{3}}{2}-36g^{4}C_A\mmagb^{4}+72g^{6}\mmagb^{6}\right.\nonumber \\
 & + & \left.12\pi^{2}\mmagb^{2}\left(\frac{g^{2}C_A^{2}}{2\pi^{2}}+\frac{8C_A+14T_F}{5}+48T_F\mubar^{2}+96T_F\mubar^{4}\right)\right\},
\ea
with $\mmagb\equiv\mmag/(g^{2}T)=c_f \frac{3}{32}C_A$ (with $c_f \geq1$) and $\mubar\equiv\mu/(2\pi T)$ as before.
In our numerical implementation, we included terms up to $p_\rmi{high,10}^\rmi{E}$ and used a cut-off $\Lambda\sim 160 \sqrt{T^2 + \mu^2}$, for which the error in the sum of $p_\rmi{1}^\rmi{E}$
and $p_\rmi{2}^\rmi{E}$ due to the introduction of the cut-off and the use of a high-momentum expansion in the second part is negligible.

\subsubsection{IR mass in Euclidean space} \label{sec:irmassine}

The IR regulating mass $m$ only appears in Eq.~(\ref{eq:pringfinite}) in two terms that are independent of the self energies, and thus the difference of
$p_\rmi{ring}^\rmi{finite}$ evaluated with masses $m_0$ and $m$ is given by
\begin{equation}
p_\rmi{ring}^\rmi{finite}(m)-p_\rmi{ring}^\rmi{finite}(m_{0})=-\frac{d_{A}C_{A}^{2}g^{4}T^{4}}{48}\sumint_{P}\left\{ \frac{1}{\left(P^{2}+m^{2}\right)^{2}}-\frac{1}{
\left(P^{2}+m_{0}^{2}\right)^{2}}\right\}. \label{pringfindiff}
\end{equation}
From Eq.~(\ref{euclmass1}), we see that the Euclidean contribution to this sum-integral is
\begin{equation}
p_{{\rmi {ring}}}^{{\rmi {fin,E}}}(m)-p_{{\rmi {ring}}}^{{\rmi {fin,E}}}(m_{0})=\frac{d_{A}C_{A}^{2}g^{4}T^{4}}{24(4\pi)^{2}}\log \fr{m}{m_{0}}\label{contrib_Eucl}
\end{equation}
which implies that the cancelation of $m$ in the sum of the Euclidean contributions to $p_\rmi{ring}^\rmi{safe}$ can be tested independently
of the Minkowskian ones by verifying that the $m$ dependence of the numerical result for $p_\rmi{1}^\rmi{E}+p_\rmi{2}^\rmi{E}$ is exactly of the above form. We have done
so with the expected positive result.

\subsection{Minkowski space contributions}

\subsubsection{Contour deformation} \label{sed:contourdeformation}

\begin{figure}\label{contdef}
\begin{center}\includegraphics[%
  scale=0.8]{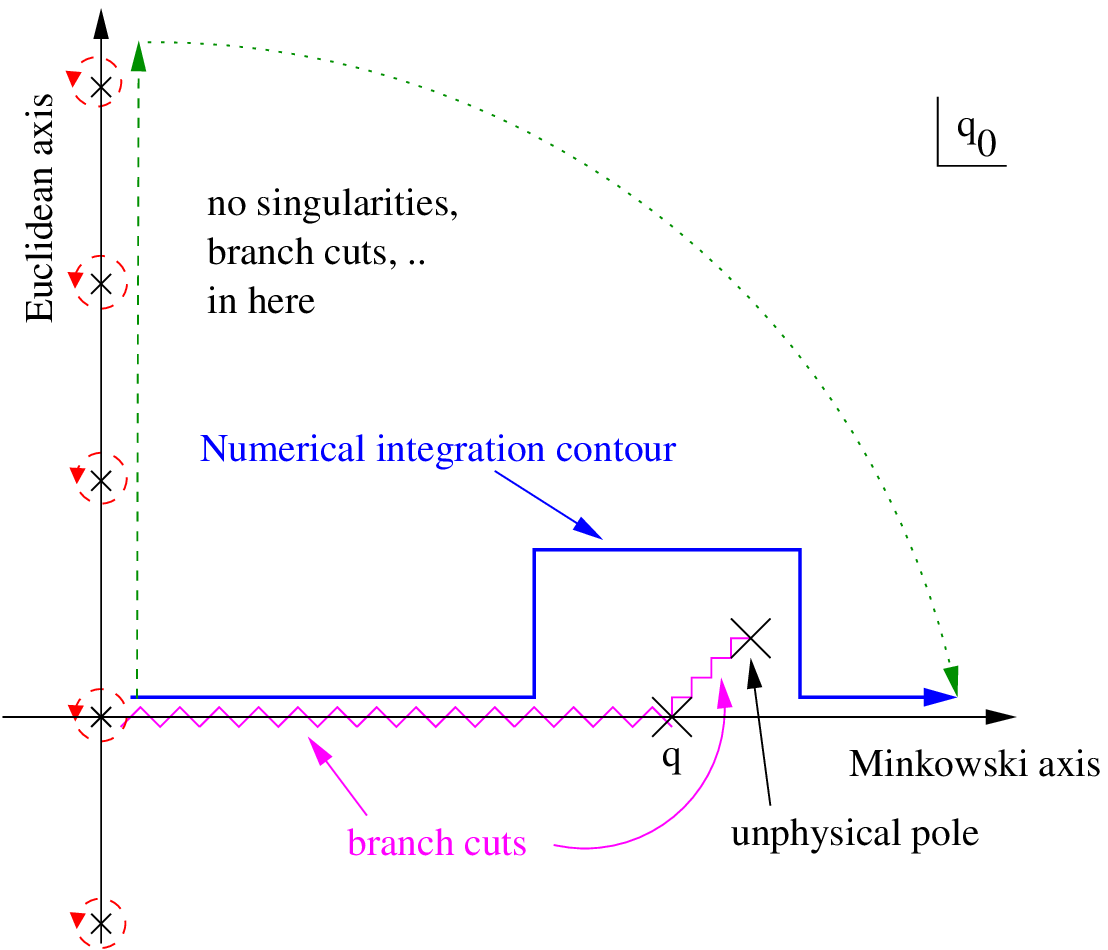}\end{center}

\caption{Symbolic illustration of the analytic structure of the integrand of $p_\rmi{ring}^\rmi{finite}$ on the complex
plane. The unphysical pole is avoided by using the complex integration path described in the text. \label{fig:analyticstructure}}
\end{figure}

As discussed in Sec.~\ref{sec:numiss}, even after the introduction 
of the magnetic mass in $p_\rmi{ring}^\rmi{finite}$ there is an
unphysical pole remaining on the complex $q_0$ plane which prohibits 
the use of the preferred ``$n_b$'' integration path along the Minkowskian axis.
Instead, we have to choose our integration contour so that we avoid the unphysical pole, illustrated in
Fig.~\ref{fig:analyticstructure}.
Since a standard path along the Minkowskian axis would fail at the branch cut from $q_0=q$
to the unphysical pole, our numerical integration contour goes around the
problematic zones at a safe distance, so that (given
the vanishing great arc contribution) the final integration is equivalent
to integrating along the Euclidean axis --- or just summing the discrete Matsubara
frequencies we started from. The validity of this procedure can be checked afterwards by varying the
safety distance between our integration path and the branch cuts.

\begin{figure}
\begin{center}\includegraphics[%
  scale=0.8]{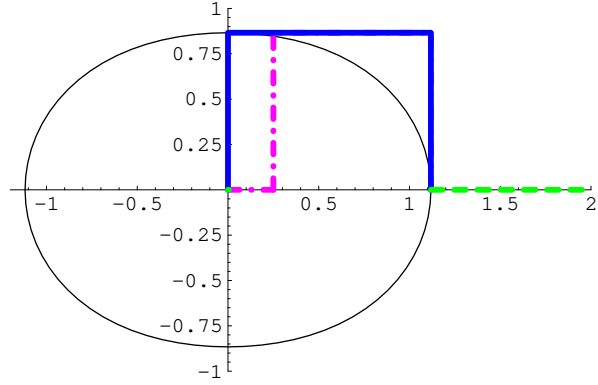}\end{center}
\caption{The shape of the Debye mass cutoff and the numerical integration contour chosen
on the complex $q_{0}$ plane, in units of $r$. The full line shows the deformed integration
path along complex $q_{0}$ while the dotted line indicates usual
Minkowski space integration, extending to $q_0\rightarrow \infty$.
The dash-dotted line shows the integration path if it is restricted
to stay in Minkowski space up to $q_{0}=q/2$. The parameters in this
and the following two Figures are chosen (a) for the big circle shape
($q=.5r$)\label{fig:shape1}}
\end{figure}

\begin{figure}
\begin{center}\includegraphics[%
  scale=0.8]{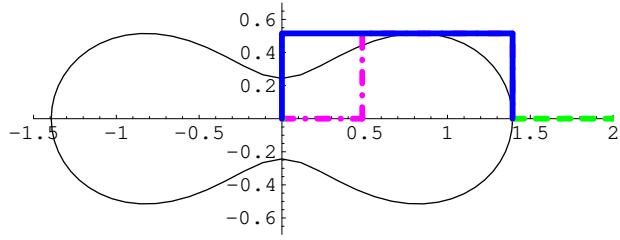}\end{center}
\caption{(b) for the eyeglasses shape ($q=.97r$)\label{fig:shape2}}
\end{figure}

\begin{figure}
\begin{center}\includegraphics[%
  scale=0.8]{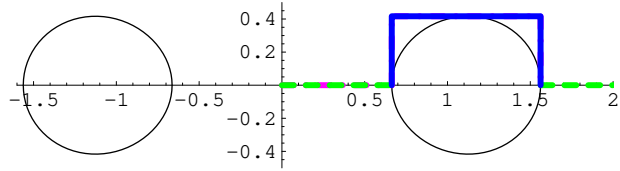}\end{center}
\caption{and (c) for the two circles shape ($q=1.2r$).\label{fig:shape3}}
\end{figure}

The unphysical poles are expected to appear in the vicinity of the lightcone. We thus want our integration path to avoid the region
$|q_{0}^{2}-q^{2}|\leq r^{2}$ where $r$ is chosen to be of the order of
the Debye mass $r=f_{c}m_\rmi{D}$ and $f_{c}>1$ is an arbitrary cutoff factor on which the final result should not depend.
We will, however, not integrate exactly along this path, but instead choose a rectangular encasing boundary.
To determine the shape of this optimal rectangular path, we write the condition
\begin{equation}
|q_{0}^{2}-q^{2}|=|(a+ib)^{2}-q^{2}|<r^{2}
\end{equation}
in terms of the real quantities $a=\re \,q_{0}$ and $b=\im \,q_{0}$ which gives the relation
\ba
4a^{2}b^{2}+(a^{2}-b^{2}-q^{2})^{2}<r^{4}.
\ea
Solving for the extremal points for $a$ and $b$ from here and denoting $x\equiv\sqrt{r^{2}+q^{2}}$,
$y\equiv\sqrt{|r^{2}-q^{2}|}$, $z\equiv r^{2}/(2q)$, we find the three cases
\begin{eqnarray*}
q<\frac{r}{\sqrt{2}}: & q_{0}\in\{0,iy,x+iy,x\} & {\rm (big\,\, circle)}\\
\frac{r}{\sqrt{2}}\leq q<r: & q_{0}\in\{0,iz,x+iz,x\} & {\rm (eyeglasses)}\\
q\geq r: & q_{0}\in\{ y,y+iz,x+iz,x\} & {\rm (two\,\, circles)}
\end{eqnarray*}
where the notation $q_{0}\in\{u,v,w,z\}$ implies that the integration path on the complex $q_0$ plane proceeds along straight lines through these points.
In practise, it is convenient to start the integration
along the Minkowskian axis also in the two first cases, so from 0 to $q/2$ the integration is always performed in Minkowski space.
The three contours are depicted in Figs.~\ref{fig:shape1} -- \ref{fig:shape3}.

With the above integration paths, the notation in Eq.~(\ref{eq:pfinringv2}) becomes obvious. The pieces $p_\rmi{L}^\rmi{M}$ and $p_\rmi{T}^\rmi{M}$ denote the
integrations performed along the Minkowskian axis (dashed lines), making use of optimizations \cite{moore}, while $p_\rmi{LD}^\rmi{M}$ and $p_\rmi{TD}^\rmi{M}$
are computed using the deformed paths (full lines or dash-dotted lines) for complex $q_{0}$.
The Minkowskian contributions $p_\rmi{L}^\rmi{M}$ and $p_\rmi{T}^\rmi{M}$ are calculated as
\begin{equation}
p_\rmi{L/T}^\rmi{M}=-\frac{d_A}{2\pi^{3}}\int_{0}^{\infty}dq\, q^{2}\int_{q_{0,\rmi{min}}}^{\infty}dq_0\,\im(p_\rmi{int}^\rmi{M}(q_0,q))\label{eq:pM}
\end{equation}
where $q_{0,\rmi{min}}$ denotes the starting point of the green
dashed lines in Figs.~\ref{fig:shape1} -- \ref{fig:shape3} (with the exception of the two circles shape for which the integration contour is composed of two disjoint pieces)
and $p_\rmi{int}^\rmi{M}$ is the
straightforward analytic continuation of Eq.~(\ref{eq:pE3}) from Euclidean to Minkowski space.
In Minkowski space, the Boltzmann factor
$n_b$ provides a natural UV cutoff and the IR mass dependent term can be evaluated separately (see below). 

The contribution of the deformed paths, $p_\rmi{LD}^\rmi{M}+p_\rmi{TD}^\rmi{M}$, is computed similarly as
\begin{equation}
p_\rmi{LD/TD}^\rmi{M}=-\frac{d_A}{2\pi^{3}}\int_{0}^{\infty}dq\, q^{2}\int_\rmi{Detour-path}dq_{0}\,\im(p_\rmi{int}^\rmi{M}(q_{0},q)).
\label{eq:pMDetour}
\end{equation}

\subsubsection{IR mass in Minkowski space\label{sub:minkowksimasscorrection}}

The so far unevaluated last contribution to Eq.~(\ref{eq:pfinringv2}) is $p_\rmi{$m$}^\rmi{M}$ which is the IR mass correction term corresponding to the $m$-dependent
terms of Eq.~(\ref{eq:pringfinite}). This integral can be computed separately from the other Minkowskian pieces, since the bosonic
distribution function provides a natural UV cutoff
for the Minkowski space calculations, making the subtraction of the quartic term from $p_\rmi{ring}^\rmi{finite}$ unnecessary.
Writing $p_\rmi{$m$}^\rmi{M}$ down explicitly (see the discussion below Eq.~(\ref{eq:pE3}) on how to include $m_\rmi{mag}$), we obtain
\begin{equation}
p_\rmi{$m$}^\rmi{M} = \frac{d_{A}C_{A}^{2}g^{4}T^{4}}{24}\int_{0}^{\infty}\frac{dq_0}{\pi}\int\frac{d^3q}{(2\pi)^{3}}n_{b}\,{\rm Im} \frac{1}{\left(Q^{2}-m^{2}\right)^{2}}\label{eq:pmM}
\end{equation}
which, not surprisingly, is observed to exactly cancel the third term of Eq.~(\ref{f}).
Thus we see that all dependence on the regulator mass $m$ cancels between the different Minkowskian contributions to $p_\rmi{ring}^\rmi{safe}$ which
together with our observation after Eq.~(\ref{contrib_Eucl}) explicitly verifies the
independence of our final result $p_\rmi{IV}$ of $m$.


\end{document}